\documentclass{article}
\usepackage[utf8]{inputenc}
\usepackage{mathtools}
\usepackage{authblk}
\usepackage{parskip}
\usepackage{multirow}
\usepackage[toc,page]{appendix}
\usepackage{comment}
\usepackage[sorting=none]{biblatex}
\usepackage{graphicx}
\usepackage{epsfig}
\usepackage{amsmath}
\usepackage{subcaption}
\graphicspath{ {./image/} }
\usepackage{amssymb}
\usepackage{amsfonts}
\usepackage{float}
\providecommand{\keywords}[1]
{
  \small	
  \textbf{\textit{Keywords---}} #1
}
\numberwithin{equation}{section}
\usepackage{hyperref}
\hypersetup{
    colorlinks=true,
    linkcolor=blue,
    filecolor=magenta,      
    urlcolor=cyan,
    pdftitle={Overleaf Example},
    pdfpagemode=FullScreen,
    }
\usepackage{geometry}
\begin{document}
\date{}
\title{Black Holes in 4D Einstein-Gauss-Bonnet gravity with  exponential electrodynamics}

\author[1]{Prosenjit Paul\thanks{prosenjitpaul629@gmail.com}}
\affil[1]{Indian Institute Of Engineering Science and Technology (IIEST), Shibpur-711103, WB, India}
\maketitle

\begin{abstract}
We investigate $4D$ Einstein-Gauss-bonnet gravity coupled to exponential electrodynamics in AdS background and found a static, spherically symmetric exact black hole solution. The horizon structure of black hole is discussed. Treating the cosmological constant as pressure, we drive first law of black hole thermodynamics in extended phase space. We compute the electrostatic potential and vacuum polarization of the black hole. To study local thermodynamical stability specific heat of the black hole is computed. Finally, we study Helmholtz free energy to analyse the global stability of black holes.
\end{abstract}
\newpage

\section{Introduction}\label{sec:1}
General Relativity(GR)  is one of the most powerful theories of classical gravitation. It precisely describes our present understanding of the observable universe, from cosmology and black holes to two-body problems, the precession of Mercury. Many other astrophysical phenomena have been verified by observational and experimental tests. However, GR is not a complete theory and the question, of how to incorporate gravity with quantum mechanics remains open. In past, people modify GR by adding higher order term to the action of GR. Here we provide some examples of modified theories of gravity,  scalar-tensor theories of gravity \cite{Barrabes:1997kk,Cai:1996pj,Capozziello:2005bu,Sotiriou:2006hs,Moffat:2005si,Faraoni:2007yn},  Lovelock theory of gravity \cite{Lovelock:1971yv,Lovelock:1972vz,Deruelle:1989fj,Konoplya:2020qqh} and brane world cosmology \cite{Cline:2000xn,Nihei:2004xv,Demetrian:2005sr}. 

In low-energy string theories, the model of gravity contains higher-order terms. Lovelock theories of gravity \cite{Lovelock:1971yv,Lovelock:1972vz} state that, assuming the spacetime dimension is four, diffeomorphism invariance, metricity and second-order equations of motion, then General Relativity with a cosmological constant is the unique theory of gravity \cite{Glavan:2019inb}. Gauss-Bonnet theory of gravity can be obtained from Lovelock's theories in higher dimensions \cite{Lanczos:1938sf}. In four dimensions Gauss-Bonnet term does not contribute to the dynamics of the theory. However, when spacetime dimensions are greater than four it contributes to the dynamics of the theory. In recent days, Glavan $\&$ Lin \cite{glavan2020einstein} obtained a $4-$dimensional Einstein-Gauss-Bonnet(EGB) theory of gravity by rescaling Gauss-Bonnet coupling parameter $\alpha$ by $\alpha/D-4$, where $D$ is the spacetime dimensions and a new static, spherically symmetric black hole solution are obtained. This black hole solution was obtained in different contexts, e.g. quantum correction to gravity \cite{Cai:2009ua,Cai:2014jea,Cognola:2013fva,Tomozawa:2011gp}. The solution of Einstein-Gauss-Bonnet theory in Maxwell electrodynamics with a negative cosmological constant was found in Ref. \cite{Fernandes:2020rpa}. In the limit $\alpha \to 0$ this solution reduces to Reissner–Nordstrom AdS black hole solution of General Relativity. In $D \to 4$ Einstein-Gauss-Bonnet was obtained in \cite{Hennigar:2020lsl} generalizing the method of Mann and Ross \cite{Mann:1992ar}. Derivation of Regularized field equations in $4D$ Einstein-Gauss-Bonnet Theory discussed in Ref. \cite{Fernandes:2020nbq}. For a complete review of $4D$ Gauss-Bonnet gravity, see Refs. \cite{Glavan:2019inb,Fernandes:2022zrq,Mahapatra:2020rds,Fernandes:2021ysi}. Thermodynamics and phase transition of $4D$ EGB black holes in AdS background studied in Refs. \cite{Chatterjee:2013daa,Hegde:2020xlv,Wei:2020poh,Wang:2020pmb,Konoplya:2020qqh,,Marks:2021fpe,Hegde:2020yrd,EslamPanah:2020hoj}. Thermodynamics of $3D$ Gauss-Bonnet black holes studied in Ref. \cite{Hennigar:2020drx}. Thermodynamics of Clouds of strings and Bardeen black holes in $4D$ Einstein-Gauss-Bonnet studied in Refs. \cite{Singh:2020nwo,Singh:2020xju,Singh:2021xbk,Kumar:2020uyz}. The rotating black hole in $4D$ Einstein-Gauss-Bonnet are also investigated in Refs. \cite{Kumar:2020owy,Papnoi:2021rvw,Zahid:2022eeq,Gammon:2022bfu}.

Born-Infeld (BI) electrodynamics \cite{Born:1934gh} is a special theory, as BI electrodynamics appear in low energy string theory \cite{Tseytlin:1986ti,Seiberg:1999vs,Gitman:2014fea}. The Lagrangian of BI electrodynamics is given by
\begin{equation}\label{eq:1.1}
    \mathcal{L}=4 \beta^2 \biggl[ 1- \sqrt{1+ \frac{\mathcal{F}}{2 \beta^2}} \biggl].
\end{equation}
In the limit $\beta \to \infty$ above Lagrangian reduces to Maxwell Lagrangian
\begin{equation}\label{eq:1.2}
    \mathcal{L}=-\mathcal{F} +\frac{1}{\beta^2} \mathcal{O}(\mathcal{F}^2),
\end{equation}
i.e. BI lagrangian reduces to Maxwell Lagrangian in the limit $\beta \to \infty$. Einstein gravity coupled to BI electrodynamics studied in \cite{Hoffmann:1935ty,Breton:2003tk,Cai:2004eh,Hendi:2014kha,Panahiyan:2018gzv}. Thermodynamics of black hole in $4D$ EGB gravity coupled with BI electrodynamics studied in Refs.\cite{Yang:2020jno,Zhang:2021kha,Yerra:2022eov}. In \cite{Hendi:2012zz} a model of Nonlinear electrodynamics is proposed, which is known as Exponential nonlinear electrodynamics(ENLE). The Lagrangian is given by
\begin{equation}\label{eq:1.3}
    \mathcal{L}=  \beta^2 \biggr[ \exp{(-{\mathcal{F}}/{\beta^2})} -1   \biggr].
\end{equation}
In the limit $\beta \to \infty$ the exponential Lagrangian reduces to Maxwell Lagrangian
\begin{equation}\label{eq:1.4}
    \mathcal{L}=-\mathcal{F} +\frac{1}{2\beta^2} \mathcal{O}(\mathcal{F}^2).
\end{equation}
Black Holes in Einstein gravity coupled to exponential, power Maxwell invariant and Logarithmic electrodynamics studied in Refs. \cite{Hendi:2012um,Hendi:2013dwa,Hendi:2014mna,Hendi:2015ixa,Hendi:2016usw}. Black holes in $\arcsin$ and exponential electrodynamics studied in Refs. \cite{Kruglov:2015fcd,Kruglov:2016ezw,Kruglov:2019okd,Kruglov:2017fck}. Some other static and spherically symmetric black holes in Einstein's gravity coupled with nonlinear electrodynamics studied in Refs. \cite{Balart:2014cga,kruglov2015nonlinear,Kruglov:2016ymq,Bronnikov:2017xrt,Bronnikov:2017sgg,Bronnikov:2017tnz,Bronnikov:2022ofk,Barrientos:2022bzm}. Recently, rotating black holes in nonlinear electrodynamics was studied in Refs. \cite{Kubiznak:2022vft,Panotopoulos:2018rjx}. Black holes in $4D$ EGB gravity coupled with nonlinear electrodynamics studied in Refs. \cite{Hyun:2019gfz,Jusufi:2020qyw,Ghosh:2020ijh,Kruglov:2021btd,Kruglov:2021pdp,Kruglov:2021qzd,Kruglov:2021rqf}\cite{Kruglov:2021stm,Singh:2022dth,Jusufi:2022ukt,Singh:2022ycn}. Thermodynamics of higher dimension Einstein-Gauss-Bonnet black holes coupled with nonlinear electrodynamics studied in Refs. \cite{Hendi:2014lke,Hendi:2015oqa,Hendi:2017lgb}.

Motivated by all the above, we studied $4D$ Einstein-Gauss-Bonnet gravity coupled to exponential electrodynamics. In this paper, we discuss the exact black hole solution in $4D$ Einstein-Gauss-Bonnet gravity coupled with exponential electrodynamics in AdS background. In the appropriate limit, our solutions reduce to $4D$ Einstein-Gauss-Bonnet gravity in Maxwell electrodynamics. The horizon structure of the black hole is depicted in fig. \ref{fig:1} and \ref{fig:2}. The physical mass and Hawking temperature of the black hole are studied. To analyze the local and global stability of the black hole, we study the specific heat and Helmotz free energy. Furthermore, we compare the thermodynamics of $4D$ EGB gravity coupled to exponential nonlinear electrodynamics with Einstein gravity coupled to exponential nonlinear electrodynamics, $4D$ EGB coupled to Maxwell electrodynamics and Einstein gravity coupled to Maxwell electrodynamics.

The paper is organised as follows. In section \ref{sec:2} we discuss the $4D$ Einstein-Gauss-Bonnet gravity coupled with exponential electrodynamics in AdS background and find the exact black hole solution. The effects of nonlinear electrodynamics parameters on the black hole horizon are discussed. In section \ref{sec:3}, we study physical mass and Hawking temperature of the black hole. Taking the cosmological constant as pressure, Gauss-Bonnet and nonlinear electrodynamics parameters as variables, we drive first law of black hole thermodynamics. To investigate local and global stability of the black hole, specific heat and Gibbs free energy are computed. Finally, in section \ref{sec:4} we summarise our results.

\section{Einstein-Gauss-Bonnet in 4D with exponential electrodynamics}\label{sec:2}
In this section, we discuss the $4D$ Einstein-Gauss-Bonnet gravity coupled to exponential Nonlinear electrodynamics. In $D$ dimension action for Einstein-Maxwell Gauss-Bonnet gravity with a negative cosmological constant coupled exponential Nonlinear electrodynamics given by 

\begin{equation}\label{eq:2.1}
S= \frac{1}{16 \pi} \int d^{D}x \sqrt{-g} \Biggr[R -2 \Lambda + \alpha \mathcal{G} +\mathcal{L(F)}    \Biggr],
\end{equation}

where $g$ is determinant of the metric $g_{\mu \nu}$. $R$ is Ricci scalar, $\alpha$ is Gauss-Bonnet coupling parameter, $\mathcal{G}= R_{\mu \nu \rho \sigma} R^{\mu \nu \rho \sigma} -4 R_{\mu \nu } R^{\mu \nu } + R^2$ is the Gauss-Bonnet term. $R_{\mu \nu \rho \sigma}$ is  Riemann tensor, $R_{\mu \nu }$ is Ricci tensor and $\mathcal{F}=-F_{\mu \nu}F^{\mu \nu}=2(E^2-B^2)$ and $F_{\mu \nu}= \partial_{\mu}{A_{\nu}}-\partial_{\nu}{A_{\mu}}$ is Maxwell tensor. In $D=4$ dimensions the Gauss-Bonnet term does not contribute to the dynamics, so we rescale \cite{Glavan:2019inb} Gauss-Bonnet coupling parameter $\alpha \to \alpha/(D-4)$. Therefore action takes the following form
\begin{equation}\label{eq:2.2}
S= \frac{1}{16 \pi} \int d^{D}x \sqrt{-g} \Biggr[R -2 \Lambda + \frac{\alpha}{D-4} \mathcal{G} +\mathcal{L(F)}     \Biggr].
\end{equation}
Now considering the static and spherical symmetric solution
\begin{equation}\label{eq:2.3}
    ds^{2}= - e^{2A(r)} dt^2 +  e^{2B(r)} dr^2 +r^2 d{\Omega}_{D-2}^2.
\end{equation}

Variation of action \eqref{eq:2.2} with respect to $A_{\mu}$ gives electromagnetic field equation
\begin{equation}\label{eq:2.4}
   \partial_{\mu} \biggr[\sqrt{-g}\partial_{\mathcal{F}} \mathcal{L}(\mathcal{F})\Bigl)    F^{\mu \nu}  \biggr] = 0, 
\end{equation}
where $\partial_{\mathcal{F}}\mathcal{L}= \frac{d\mathcal{L}}{d\mathcal{F}}$. Here we consider the Exponential form of nonlinear electromagnetic field (ENLD) and the Lagrangian is given by
\begin{equation}\label{eq:2.5}
\mathcal{L}= \beta^2 \biggr[ \exp{(-{\mathcal{F}}/{\beta^2})} -1   \biggr].
\end{equation}
The electromagnetic field tensor only contributes to the radial electric field $F_{tr}=-F_{rt}$. Therefore, equation \eqref{eq:2.4} takes the following form
\begin{equation}\label{eq:2.6}
    \partial_{r} \Bigr[ \frac{ E r^{D-2} \exp{{(\frac{2E^2}{\beta^2})}}}{\exp{(A+B)}}\Bigr]=0.
\end{equation}
The electric field is given by
\begin{equation}\label{eq:2.7}
E=\frac{Qe^{A+B}}{r^{D-2}} \exp{\biggr[ -\frac{Lambert{(x)}}{2} \biggr]},
\end{equation}
where $x=\frac{4Q^2}{\beta^2 r^4}$. To find metric we use equation \eqref{eq:2.3}, equation \eqref{eq:2.7} into action \eqref{eq:2.2} and taking limit $D \to 4$ we obtain
\begin{equation}\label{eq:2.8}
S= \frac{\Sigma_{2}}{16 \pi} \int dt dr 2 e^{A+B}  \biggr[ r^{3} \psi \Bigl( 1+ \alpha \psi\Bigl) + \frac{r^{3}}{l^2} - \frac{\beta^2 y}{2}   \biggr]^{\prime},
\end{equation}

where prime denotes differentiation with respect to r,
\begin{equation*}
    y= \int dr \Bigl\{\exp{(2E^2/\beta^2)} -1 \Bigl\},
\end{equation*}
\begin{equation*}
    \Sigma_{2}=\frac{2{\pi}^\frac{3}{2}}{\Gamma\Bigl( {1+\frac{1}{2}}\Bigl)},
\end{equation*}
\begin{equation}\label{eq:2.9}
    \psi= r^{-2} \Bigl( 1- e^{-2B} \Bigl).
\end{equation}

From action \eqref{eq:2.8} we can obtain the solutions as
\begin{equation*}
    e^{A+B}=1,
\end{equation*}
\begin{equation}\label{eq:2.10}
  \alpha r^3 \psi^2 +r^3 \psi+\frac{r^3}{l^2} - \frac{\beta^2 y}{2}-2M=0, 
\end{equation}
where $ M$ is the integration constant and it is related to the mass of the black hole. Using equation \eqref{eq:2.7} and first equation of \eqref{eq:2.10} we obtain the electrostatic potential
\begin{equation}\label{eq:2.11}
    \Phi= -\frac{1}{10} \Bigl(4Q^2 \beta^2 L_W e^{-L_W} \Bigl)^{1/4} \biggr[ \frac{2}{9}L_W^2 {}_{1}F_{1} \Bigl( 1;\frac{13}{4};\frac{L_W}{4}\Bigl) +9 L_W+5 \biggr],
\end{equation}
where $L_W=LambertW(x)$ and ${}_{1}F_{1}$ is the confluent hypergeometric function \cite{abramowitz1964handbook} of first kind. Therefore, exact solution is 
\begin{equation}\label{eq:2.12}
e^{2A}= e^{-2B}= 1+ \frac{r^2}{2\alpha} \Biggr[1 \pm \sqrt{1+4\alpha \biggl\{ \frac{2M}{r^3} + \frac{\beta^2 y}{2r^3} - \frac{1}{l^2}   \biggl\} }  \Biggr],
\end{equation}
where $y$ takes the following form
\begin{equation}\label{eq:2.13}
y= \int dr \Bigl\{\sqrt{\frac{x}{Lambert(x)}} -1 \Bigl\} r^2.
\end{equation}
Next, we will verify the metric \eqref{eq:2.12} in the limit $\beta \to \infty$. Applying limit $\beta \to \infty$ to  equation \eqref{eq:2.13} we obtain 
\begin{equation}\label{eq:2.14}
    y=-\frac{2Q^2}{\beta^2 r} + \mathcal{O}(Q^4/r^5).
\end{equation}
Therefore, using above equation and equation \eqref{eq:2.12} the metric function of Exponential nonlinear electrodynamics reduces to Maxwell electrodynamics \cite{Fernandes:2020rpa}
\begin{equation}\label{eq:2.15}
e^{2A}= e^{-2B}= 1+ \frac{r^2}{2\alpha} \Biggr[1 \pm \sqrt{1+4\alpha \biggl\{ \frac{2M}{r^3} - \frac{Q^{2} }{r^{4}} - \frac{1}{l^2}   \biggl\} }  \Biggr].
\end{equation}
To find the exact black hole solutions in Exponential nonlinear electrodynamics, we have to evaluate integral of equation \eqref{eq:2.13}. After some cumbersome computation integral comes out as 
\begin{equation}\label{eq:2.16}
 y= -\frac{e^{-L_W/4}(Q/\beta)^{3/2}}{L_W^{3/4}}\Biggr[ \frac{4 \sqrt{2}}{15}{  L_{W}^{2} F\Bigr[ 1, \frac{9}{4};\frac{L_{W}(x)}{4}\Bigr] } + \frac{4 \sqrt{2}}{3} L_{W} - \frac{2\sqrt{2}}{3 } \Biggr]  - \frac{r^3}{3}.
\end{equation}
Therefore, the metric function of $4D$ EGB gravity coupled to ENLD is
\begin{equation}\label{eq:2.17}
e^{2A}= 1+ \frac{r^2}{2\alpha} \Biggr[1 \pm \sqrt{1+4\alpha \biggl\{ \frac{2M}{r^3} - \biggl(\frac{4}{3\sqrt{2}r^3} \Bigl( \frac{\beta^{10} Q^6 e^{-L_{W}} }{x L_{W}^{5}} \Bigl)^{1/8}  \Bigl\{ \frac{1}{5}L_{W}^2 {}_{1}F_{1} \Bigl( 1;\frac{9}{4};\frac{L_W}{4}\Bigl) +  L_{W}- \frac{1}{2} \Bigl\} +\frac{\beta^2}{6} \biggl) - \frac{1}{l^2}   \biggl\} }  \Biggr].
\end{equation}

\begin{figure}[H]
\centering
\subfloat[$\alpha=0.2$]{\includegraphics[width=.5\textwidth]{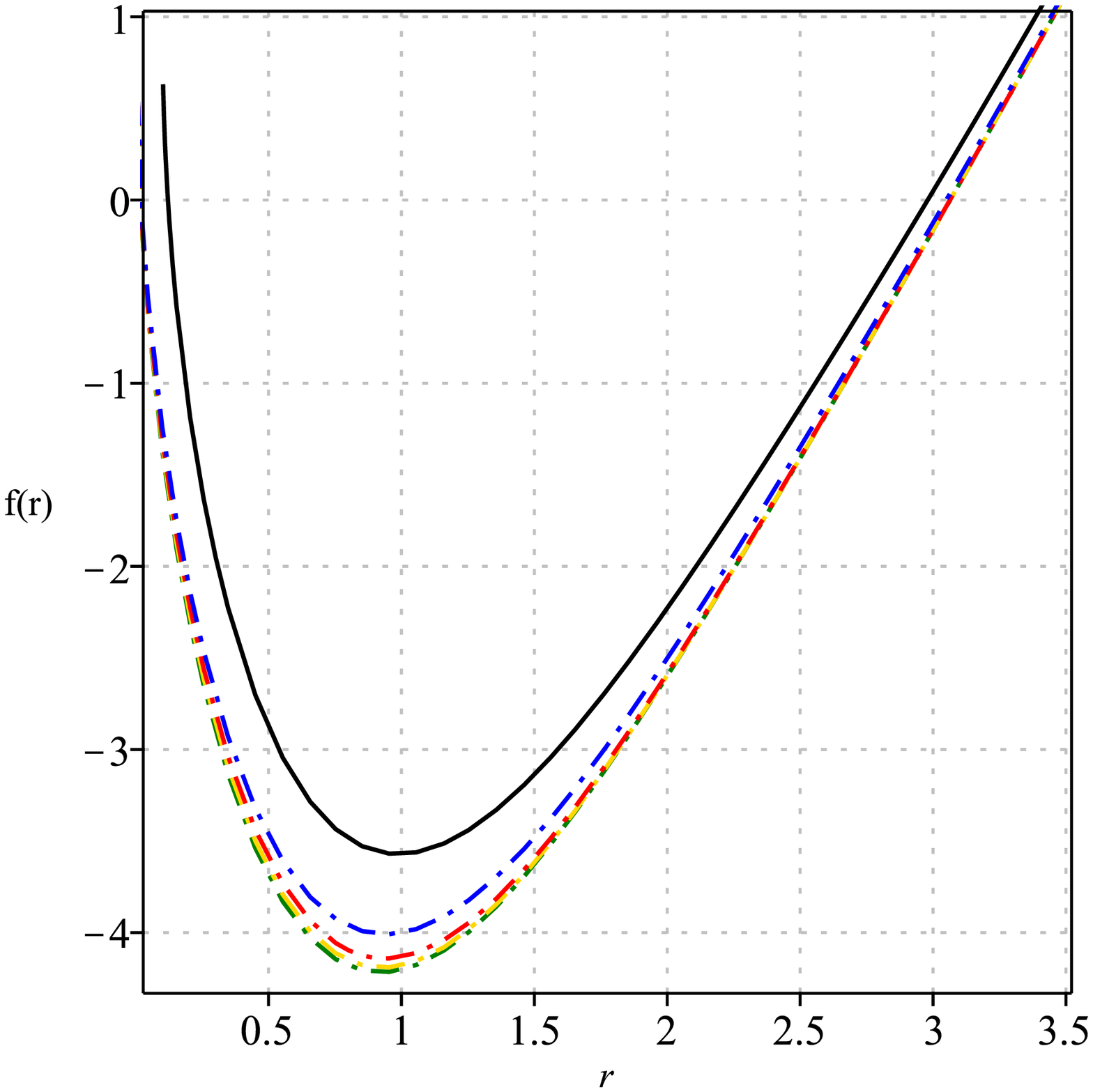}}\hfill
\subfloat[$\alpha=0.2$]{\includegraphics[width=.5\textwidth]{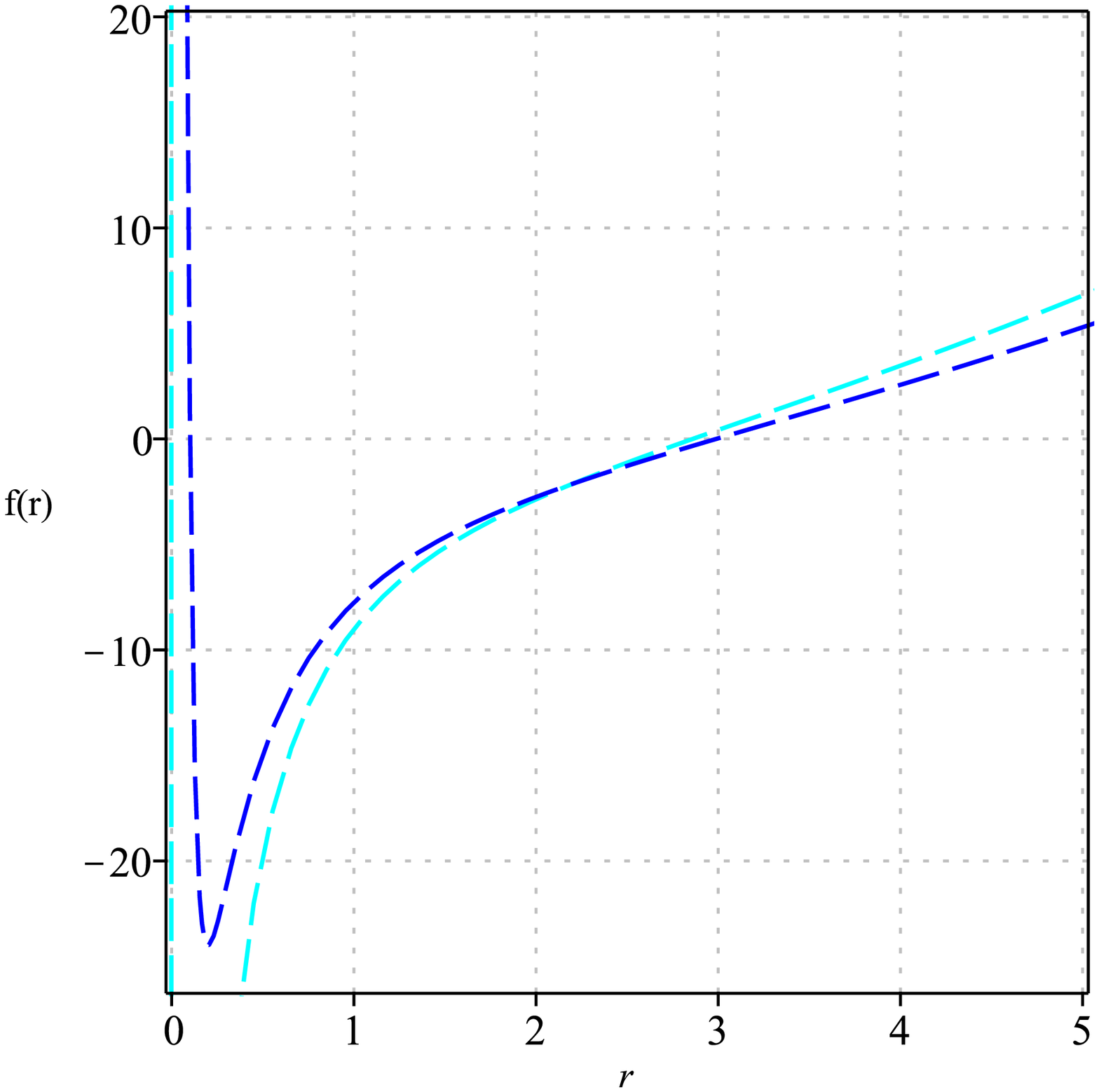}}\hfill
\caption{$\beta=0.05$ denoted by blue dash dot line, $\beta=0.5$ denoted by red dash dot line with, $\beta=1.0$ denoted by gold dash dot line and $\beta=1.5$ denoted by green dash dot line in EGB-ENLE. Solid black line denotes EGB-Maxwell, cyan dash line denotes GR-ENLE and blue dash line denotes $\beta=0.5$ in GR-Maxwell with   $Q=1$, $l=2$ and $M=5$.}\label{fig:1}
\end{figure}

\begin{figure}[H]
\centering
\subfloat[$\alpha=0.4$]{\includegraphics[width=.5\textwidth]{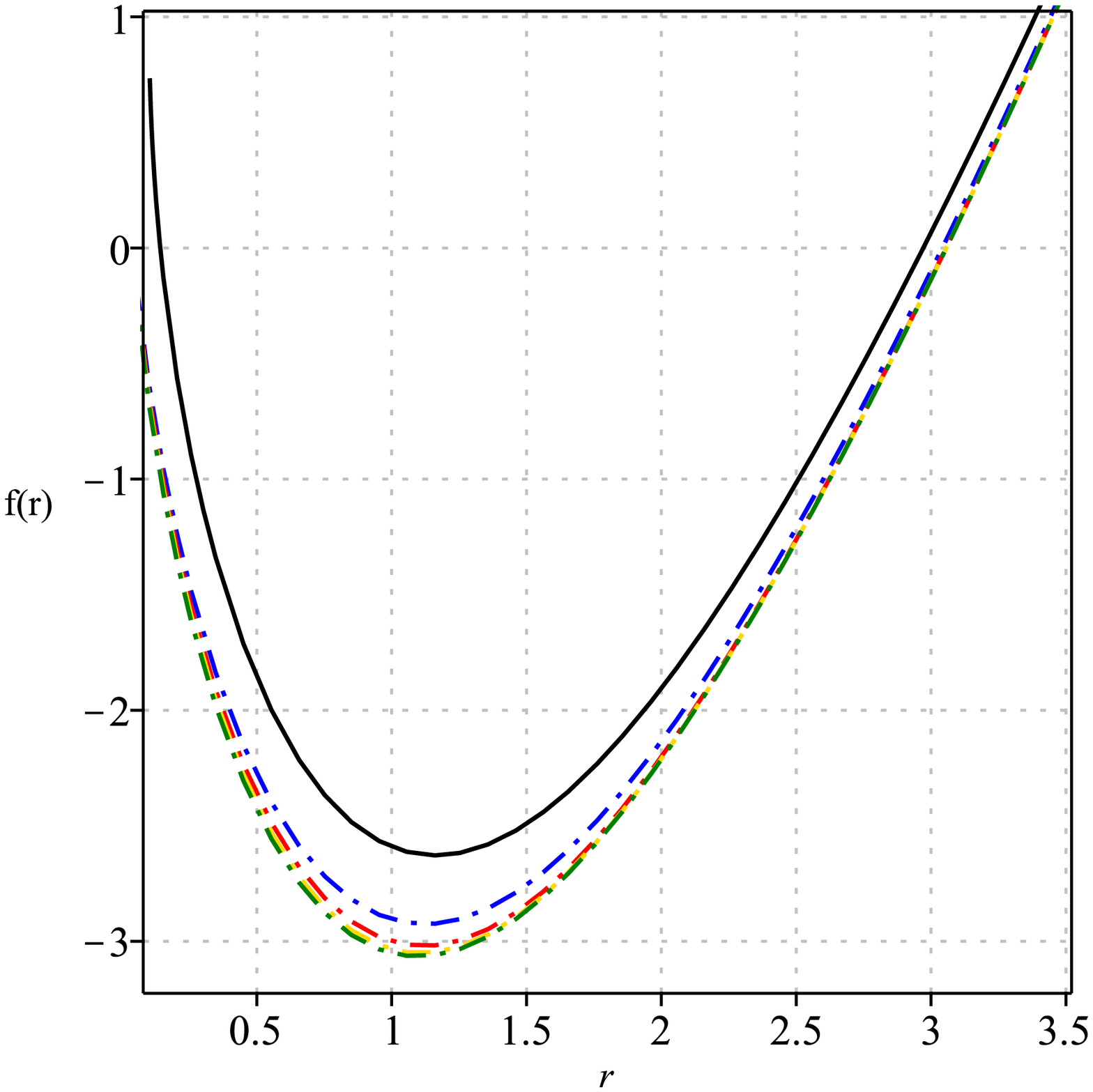}}\hfill
\subfloat[$\alpha=0.4$]{\includegraphics[width=.5\textwidth]{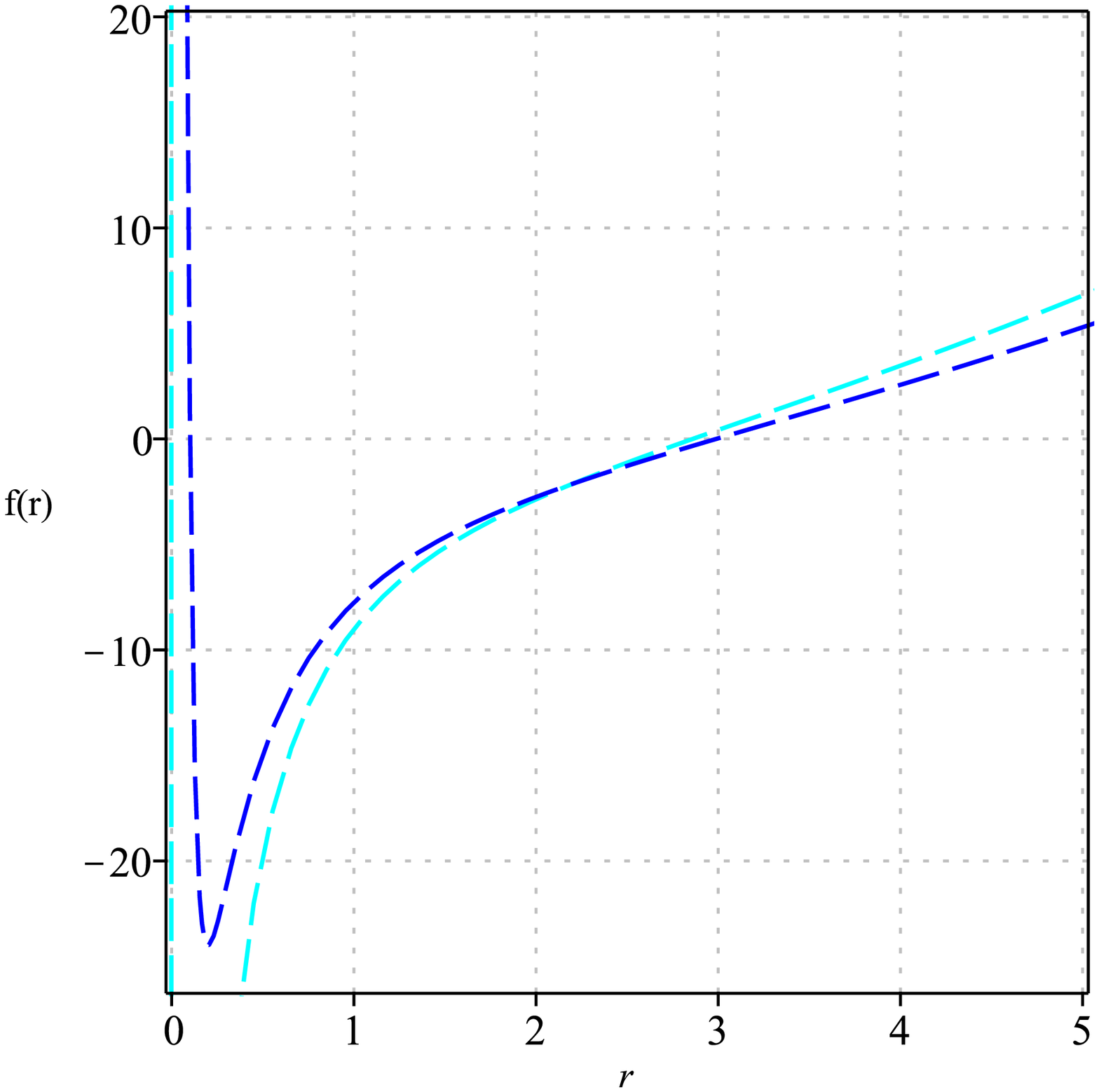}}\hfill
\caption{$\beta=0.05$ denoted by blue dash dot line, $\beta=0.5$ denoted by red dash dot line with, $\beta=1.0$ denoted by gold dash dot line and $\beta=1.5$ denoted by green dash dot line in EGB-ENLE. Solid black line denotes EGB-Maxwell, cyan dash line denotes GR-ENLE and blue dash line denotes $\beta=0.5$ in GR-Maxwell with   $Q=1$, $l=2$ and $M=5$.}\label{fig:2}
\end{figure}
In fig. \ref{fig:1} and fig. \ref{fig:2} we plot metric function of $4D$ Einstein-Gauss-Bonnet black hole coupled with exponential electrodynamics in AdS background for different values of $\beta$ and Gausss-Bonnet coupling parameters. The black hole has two horizons, if we increase the $\beta$ values then the position of outer horizon increases also. The horizon structure of $4D$ Einstein-Gauss-Bonnet black hole in Maxwell electrodynamics, black hole in Einstein gravity coupled with exponential electrodynamics and Reissner–Nordstrom AdS black hole are also depicted.     
\section{Black Hole Thermodynamics}\label{sec:3}

In this section, we study the thermodynamics of $4D$ Einstein Gauss-Bonnet gravity coupled with ENLD. The thermodynamics of $4D$ black holes in General Relativity with exponential electrodynamics in AdS background studied \cite{Hendi:2013dwa,Hendi:2014mna}. The physical mass of the black hole can be obtained from the metric function \eqref{eq:2.17} by setting $e^{-2B}\vert_{r_+}=0$
\begin{equation}\label{eq:3.1} 
M=  \Biggr[   \frac{\alpha}{2 r_+}+\frac{r}{2}+\frac{\beta^{2} r_+^{3}}{12}+\frac{r_+^{3}}{2 l^{2}} + \frac{\sqrt{2\beta Q^3}  (2 L_W-1)}{6 \sqrt{\sqrt{x} L_W} } + \frac{\beta^{2} L_W^{\frac{5}{4}} 2^{\frac{1}{4}} 4^{\frac{5}{8}}(\frac{Q^{2}}{\beta^{2}})^{\frac{3}{4}} {}_{1}F_{1} \Bigl( 1;\frac{9}{4};\frac{L_W}{4}\Bigl) { e}^{-\frac{L_W}{4}}}{30}
\Biggr]
\end{equation}
The Hawking temperature of black holes is defined as 
\begin{equation}\label{eq:3.2}
    T_H=\frac{f^{\prime}(r_+)}{4\pi}
\end{equation}
Therefore, using equation \eqref{eq:2.17} and equation \eqref{eq:3.2} we obtain
\begin{equation}\label{eq:3.3}
    T_{H}=\frac{(\beta^{2} l^{2}+6) r_+^{4}+2 l^{2} r_+^{2}-2 \alpha  l^{2}}{8 \pi  (r_+^{2}+2 \alpha ) l^{2} r}-\frac{r_+ Q \beta}{4 \pi  \sqrt{L_W} (r_+^{2}+2 \alpha )}
\end{equation}

\begin{figure}[H]
\centering
\subfloat[$\alpha=0.2$]{\includegraphics[width=.5\textwidth]{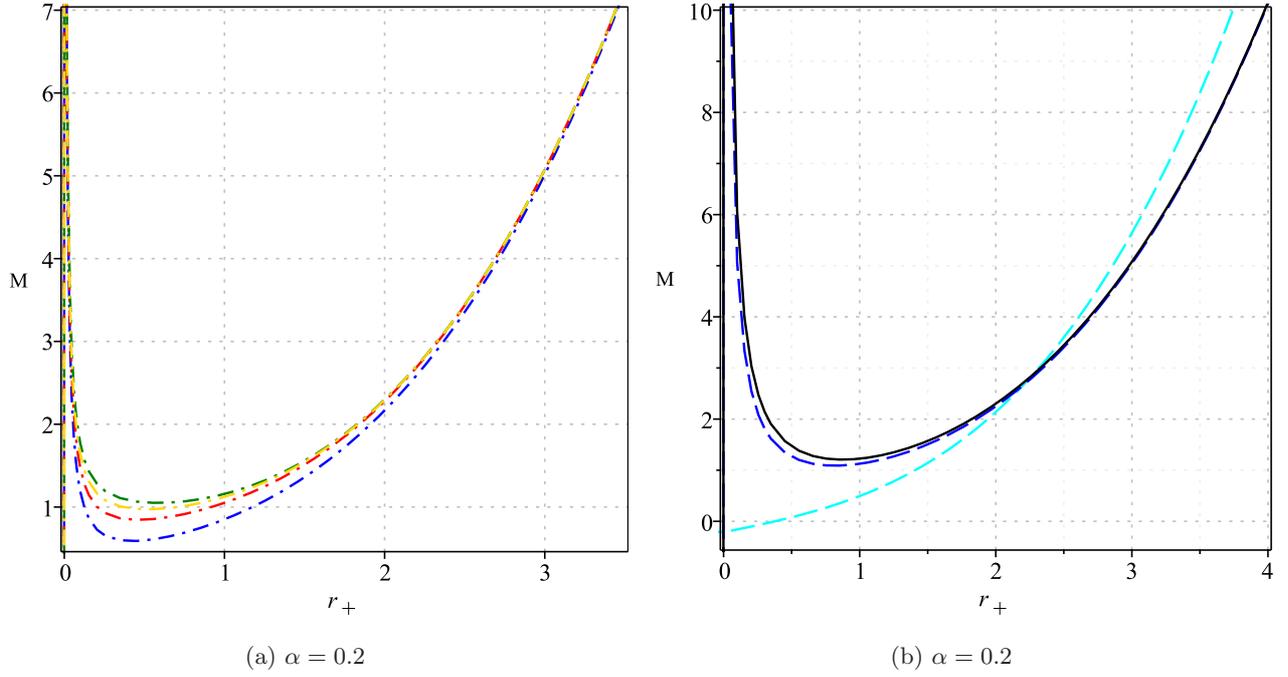}}\hfill
\subfloat[$\alpha=0.2$]{\includegraphics[width=.5\textwidth]{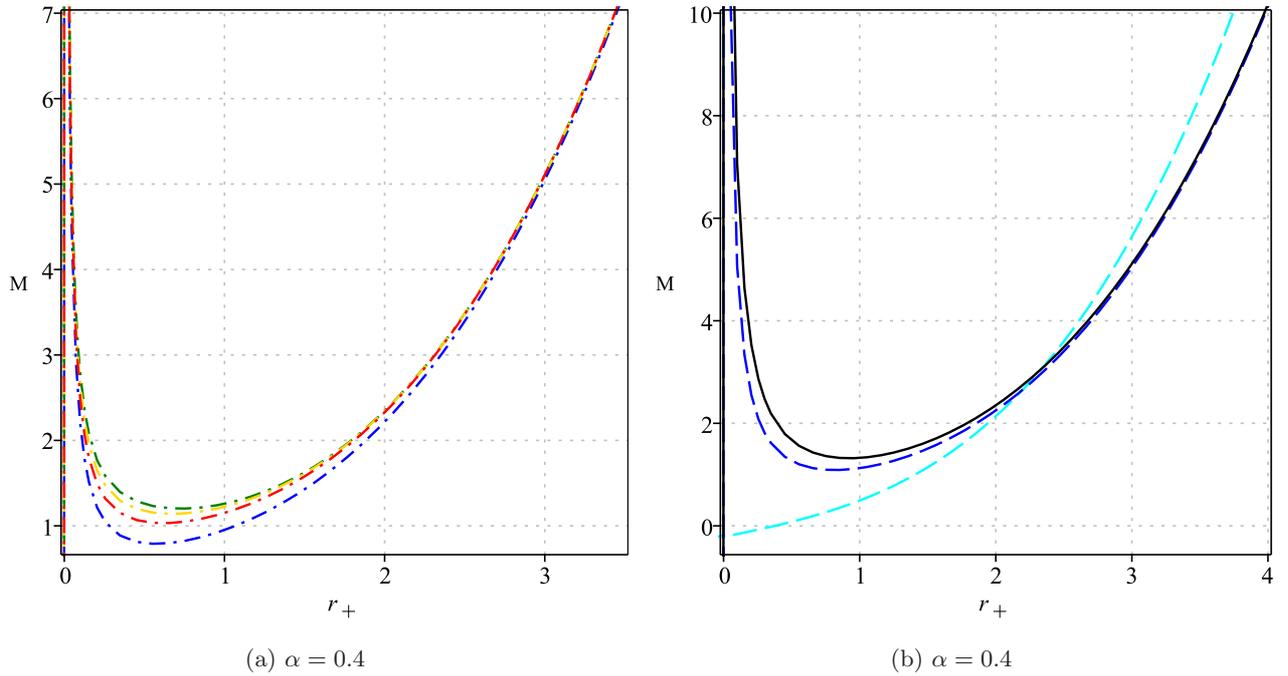}}\hfill
\caption{$\beta=0.05$ denoted by blue dash dot line, $\beta=0.5$ denoted by red dash dot line with, $\beta=1.0$ denoted by gold dash dot line and $\beta=1.5$ denoted by green dash dot line in EGB-ENLE. Solid black line denotes EGB-Maxwell, cyan dash line denotes GR-ENLE, and blue dash line denotes $\beta=0.5$ in GR-Maxwell with   $Q=1$, $l=2$ and $M=5$.}\label{fig:3}
\end{figure}

\begin{figure}[H]
\centering
\subfloat[$\alpha=0.4$]{\includegraphics[width=.5\textwidth]{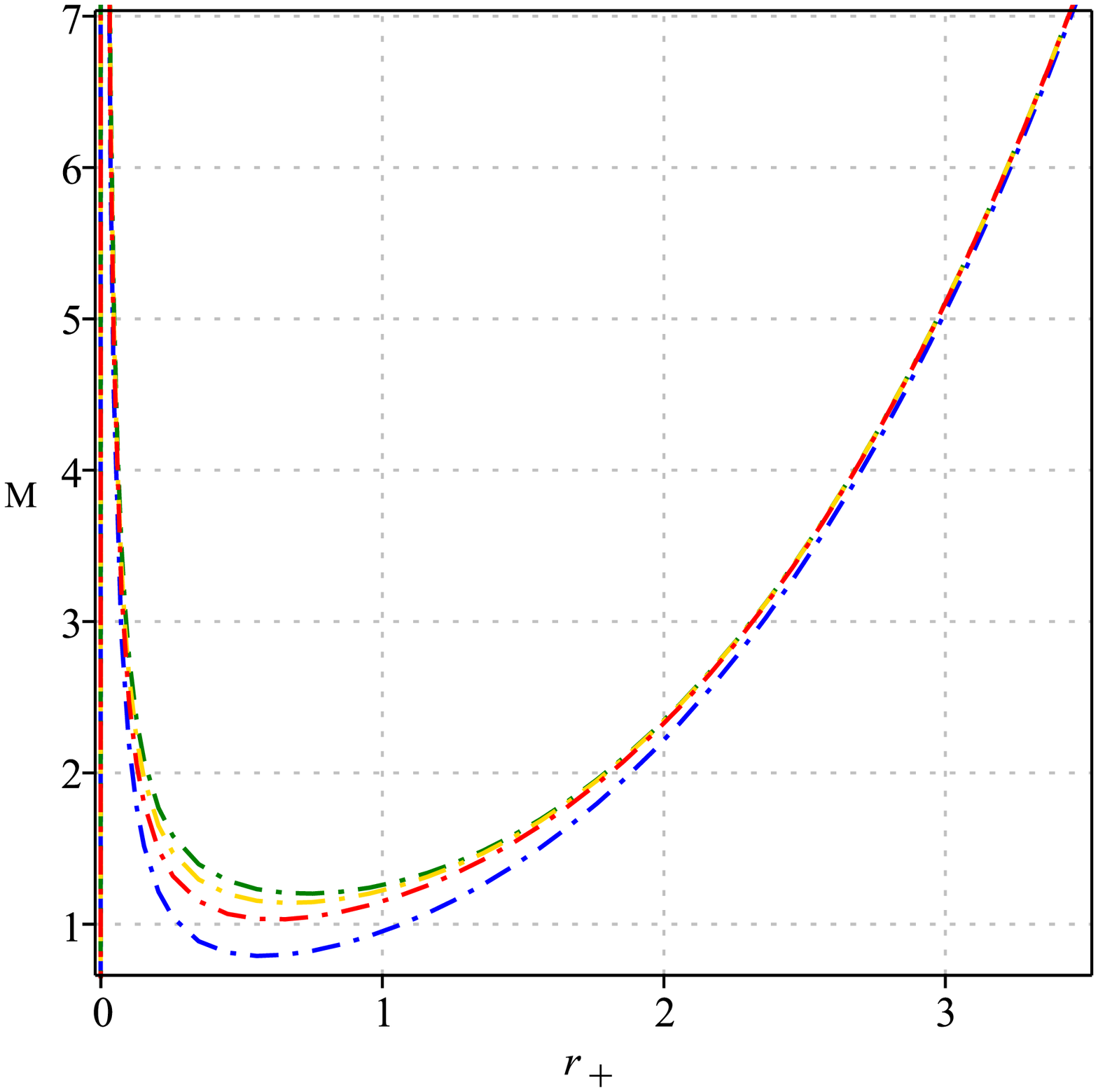}}\hfill
\subfloat[$\alpha=0.4$]{\includegraphics[width=.5\textwidth]{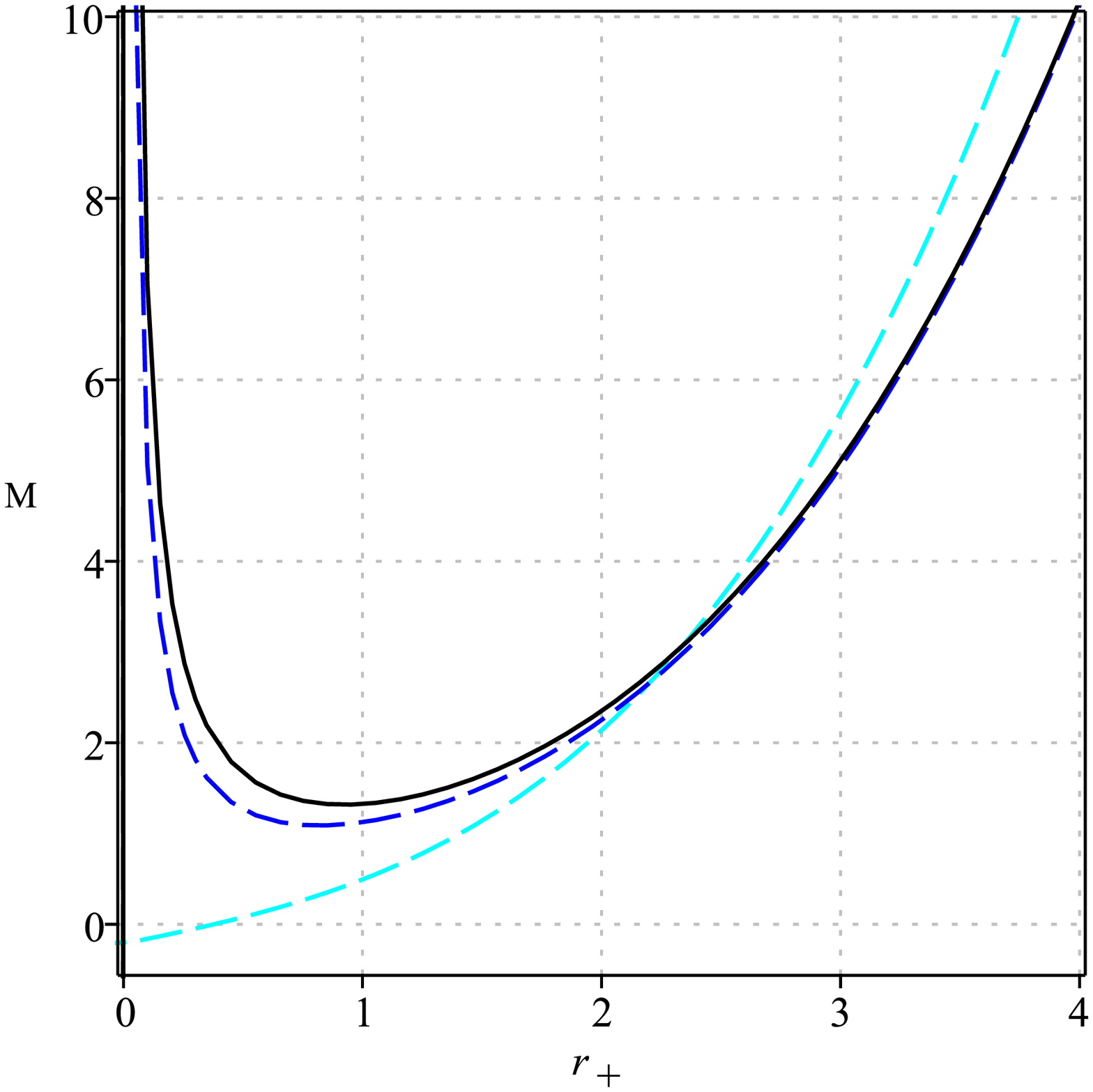}}\hfill
\caption{$\beta=0.05$ denoted by blue dash dot line, $\beta=0.5$ denoted by red dash dot line with, $\beta=1.0$ denoted by gold dash dot line and $\beta=1.5$ denoted by green dash dot line in EGB-ENLE. Solid black line denotes EGB-Maxwell, cyan dash line denotes GR-ENLE and blue dash line denotes $\beta=0.5$ in GR-Maxwell with   $Q=1$, $l=2$ and $M=5$.}\label{fig:4}
\end{figure}

The physical mass of the black hole is depicted in fig. \ref{fig:3} and fig. \ref{fig:4} for different values of $\beta$ and Gauss-Bonnet coupling parameters. For a particular value of horizon radius mass of the black hole attains a minimum and then it is an increasing function of horizon radius. If we increase the $\beta$, then minimum value of black hole mass slowly decreases. For very small-sized black hole, mass is very high. Furthermore, we plot the mass of the black hole in $4D$ EGB with Maxwell electrodynamics, Einstein gravity with ENLE and Einstein gravity with Maxwell electrodynamics. All of them is increasing function of horizon radius and for a particular value of horizon radius mass of the black hole coincide.

\begin{figure}[H]
\centering
\subfloat[$\alpha=0.2$]{\includegraphics[width=.5\textwidth]{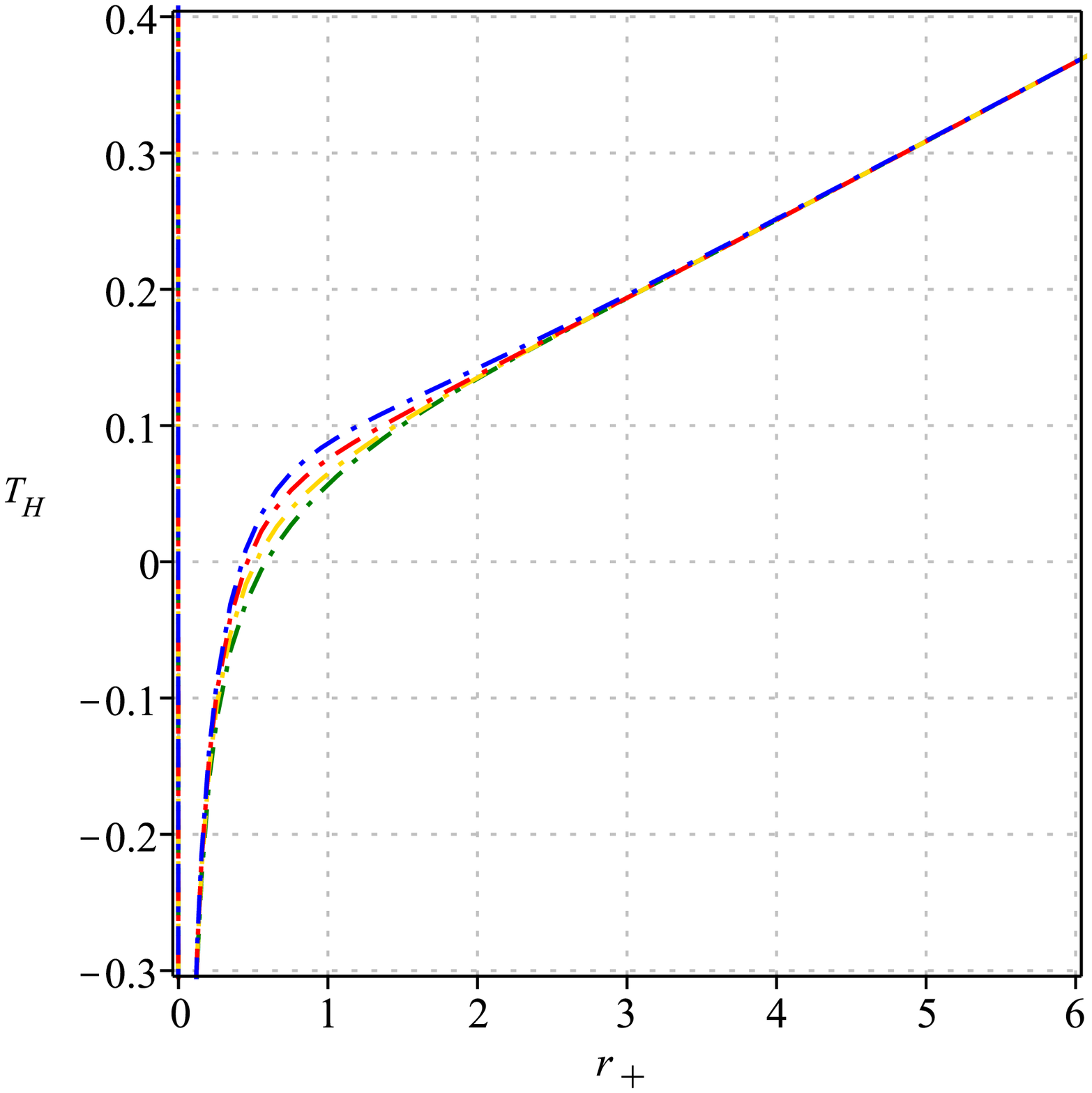}}\hfill
\subfloat[$\alpha=0.2$]{\includegraphics[width=.5\textwidth]{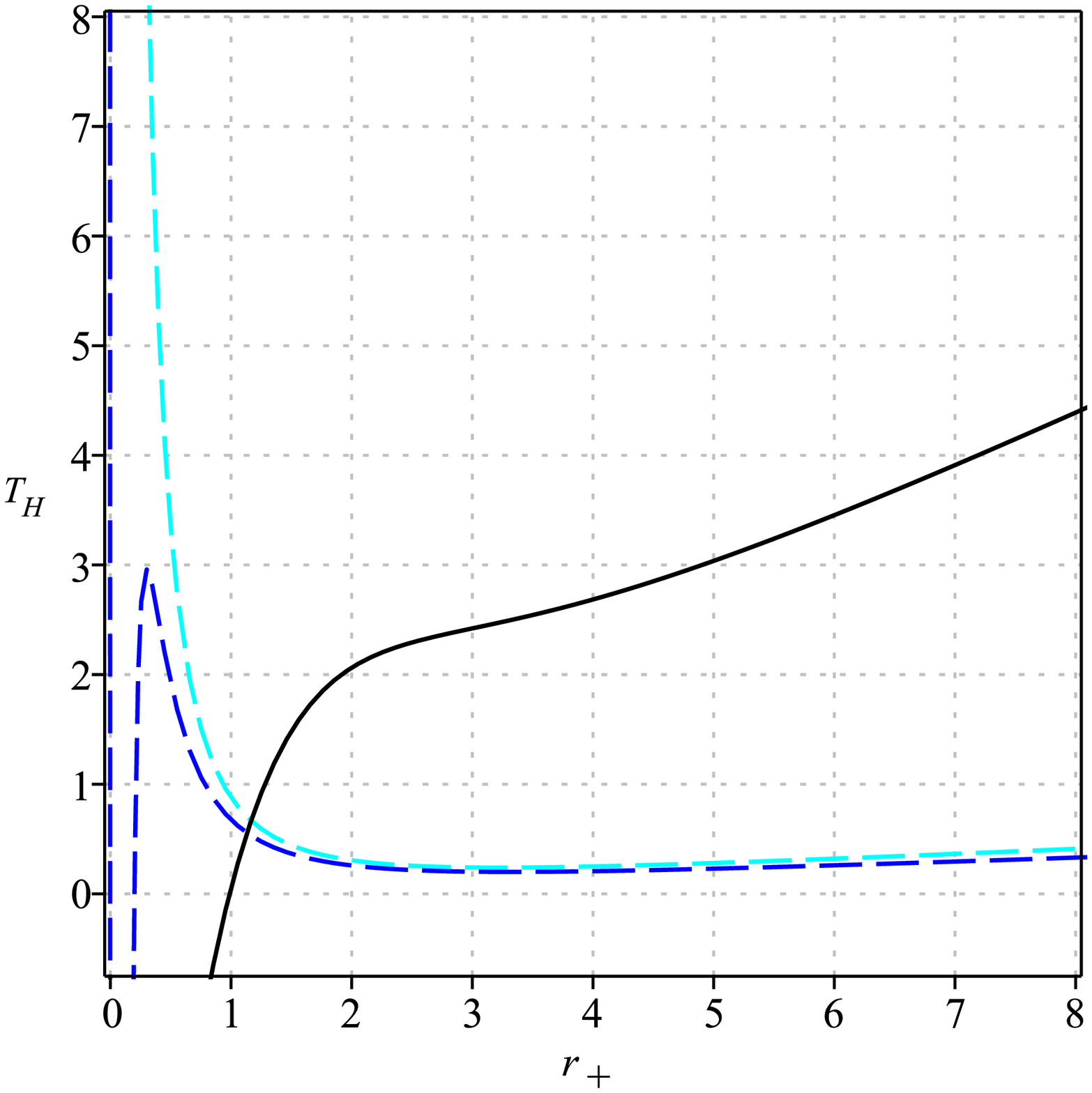}}\hfill
\caption{$\beta=0.05$ denoted by blue dash dot line, $\beta=0.5$ denoted by red dash dot line with, $\beta=1.0$ denoted by gold dash dot line and $\beta=1.5$ denoted by green dash dot line in EGB-ENLE. Solid black line denotes EGB-Maxwell, cyan dash line denotes GR-ENLE and blue dash line denotes $\beta=0.5$ in GR-Maxwell with   $Q=1$, $l=2$ and $M=5$.}\label{fig:5}
\end{figure}

\begin{figure}[H]
\centering
\subfloat[$\alpha=0.4$]{\includegraphics[width=.5\textwidth]{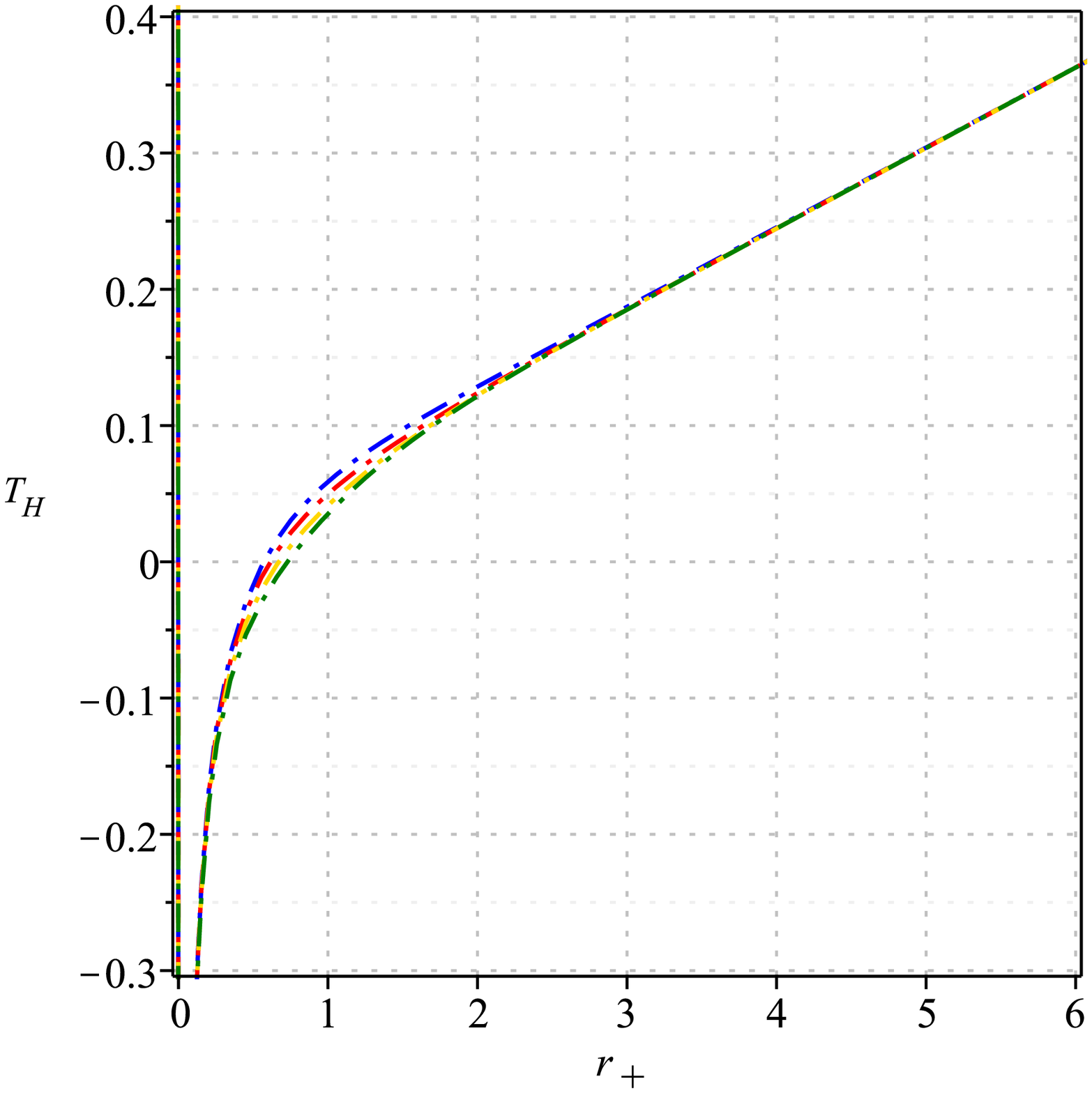}}\hfill
\subfloat[$\alpha=0.4$]{\includegraphics[width=.5\textwidth]{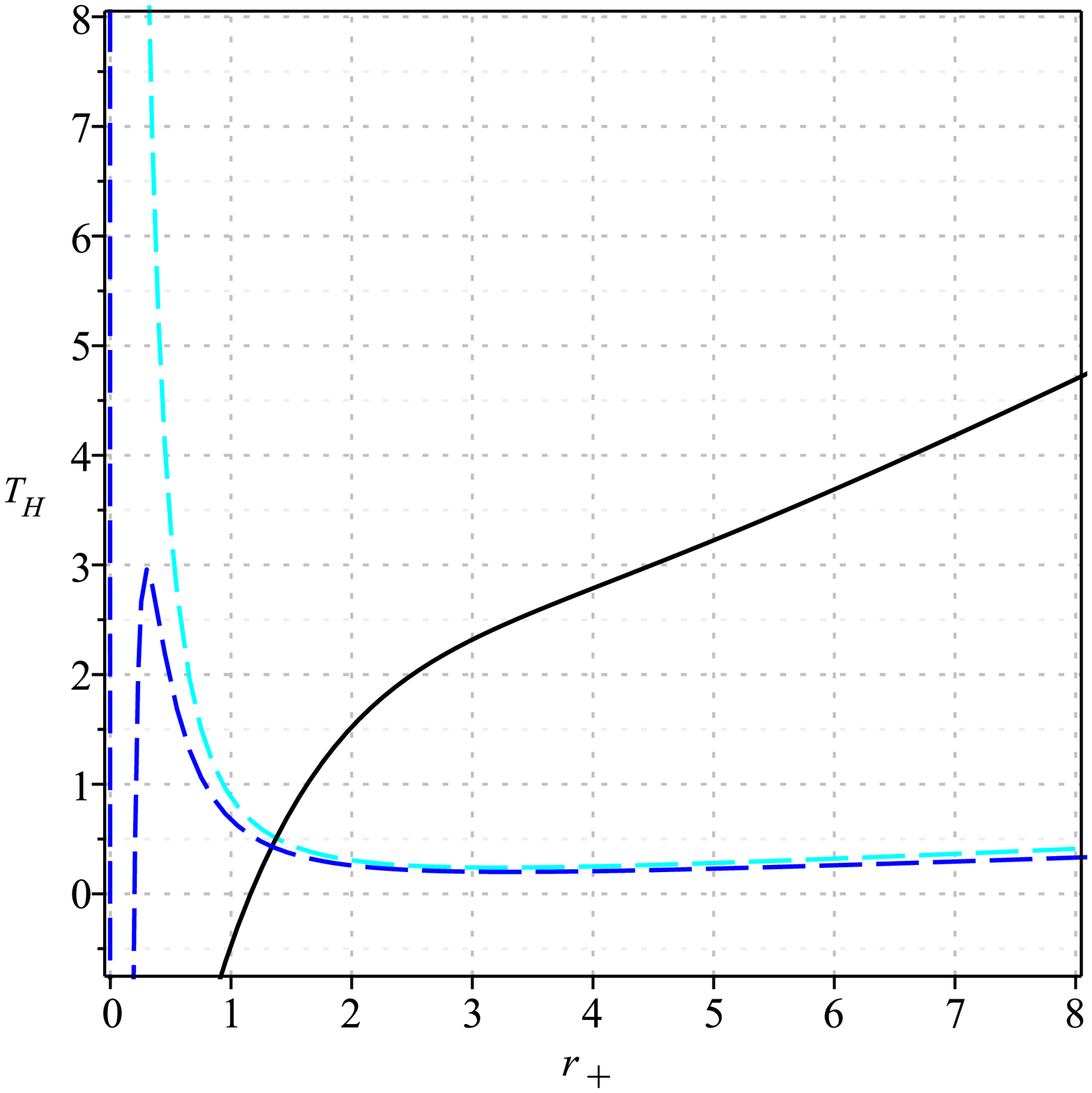}}\hfill
\caption{$\beta=0.05$ denoted by blue dash dot line, $\beta=0.5$ denoted by red dash dot line with, $\beta=1.0$ denoted by gold dash dot line and $\beta=1.5$ denoted by green dash dot line in EGB-ENLE. Solid black line denotes EGB-Maxwell, cyan dash line denotes GR-ENLE and blue dash line denotes $\beta=0.5$ in GR-Maxwell with   $Q=1$, $l=2$ and $M=5$.}\label{fig:6}
\end{figure}

The entropy of the black hole defined as 

\begin{equation}\label{eq:3.4}
    S=\int \frac{dM}{T_{H}}.
\end{equation}
Using the Hawking temperature and mass of the black hole we obtained the entropy
\begin{equation}\label{eq:3.5}
    S= \pi r_{+}^2 + 4\pi \alpha\ln(r_{+}) +S_{0},
\end{equation}

where $S_{0}$ is integration constant. To drive the first law of black hole thermodynamics we will treat the exponential nonlinear electrodynamics parameters $\beta$ as thermodynamics variables and corresponding potential is $\mathcal{B}$, which is known as vacuum polarization. Apart from that potential corresponding to Einstein-Gauss-Bonnet parameter $\alpha$ is $\mathcal{A}$. The thermodynamic pressure is defined as $P=3/8 \pi l^2$ and we will treat cosmological constant as pressure. The first law of black hole thermodynamics in extended phase space takes the following form
\begin{equation}\label{eq:3.6}
    dM= T_{H} dS + \Phi dQ +V dP + {\mathcal{A}} d\alpha   + {\mathcal{B}}  d\beta . 
\end{equation}
Now, from the first law of black hole, one can find the potential and volume as
\begin{equation*}
\mathcal{B} =\frac{1}{{270 (1+L_W ) }} \Biggr[ {4 \beta  r_+ L_W^{\frac{5}{2}} {}_{1}F_{1} \Bigl( 2;\frac{13}{4};\frac{L_W}{4}\Bigl)    -180 \sqrt{Q \beta}  (L_W+4) \Gamma  (\frac{5}{4}, \frac{L_W }{4}) }
\end{equation*}
\begin{equation}\label{eq:3.7}
+90\beta  r_+ L_W^{\frac{3}{2}}  +135 \beta  r_+ \sqrt{L_W}  +\frac{45 \pi}{\Gamma  (\frac{3}{4})}   \sqrt{2 \beta Q}  (L_W +4) \Biggr]
\end{equation}

\begin{equation}\label{eq:3.8}
    V = \biggl( \frac{\partial{M}}{\partial{P}} \biggl)_{S,Q,\alpha}= \frac{4}{3} \pi r_{+}^3,
\end{equation}
\begin{equation}\label{eq:3.9}
    \mathcal{A} = \biggl( \frac{\partial{M}}{\partial{\alpha}} \biggl)_{S,Q,P}= \frac{1}{2r_{+}} 
\end{equation}

Next, we will study the local thermodynamical stability of the black hole. we compute specific heat of the black hole. The local thermodynamical stability of the black hole can be studied from the sign of the specific heat. Black hole is thermodynamically unstable if $C_{\Phi}<0$ and for $C_{\Phi}>0$ then black hole is thermodynamically stable. The heat capacity of the black hole defined as
\begin{equation}\label{eq:3.10}
    C_{\Phi}=  \Biggl( \frac{dM}{dT_{H}}  \Biggl)_{\Phi}.
\end{equation}

\begin{figure}[H]
\centering
\subfloat[$\alpha=0.2$]{\includegraphics[width=.5\textwidth]{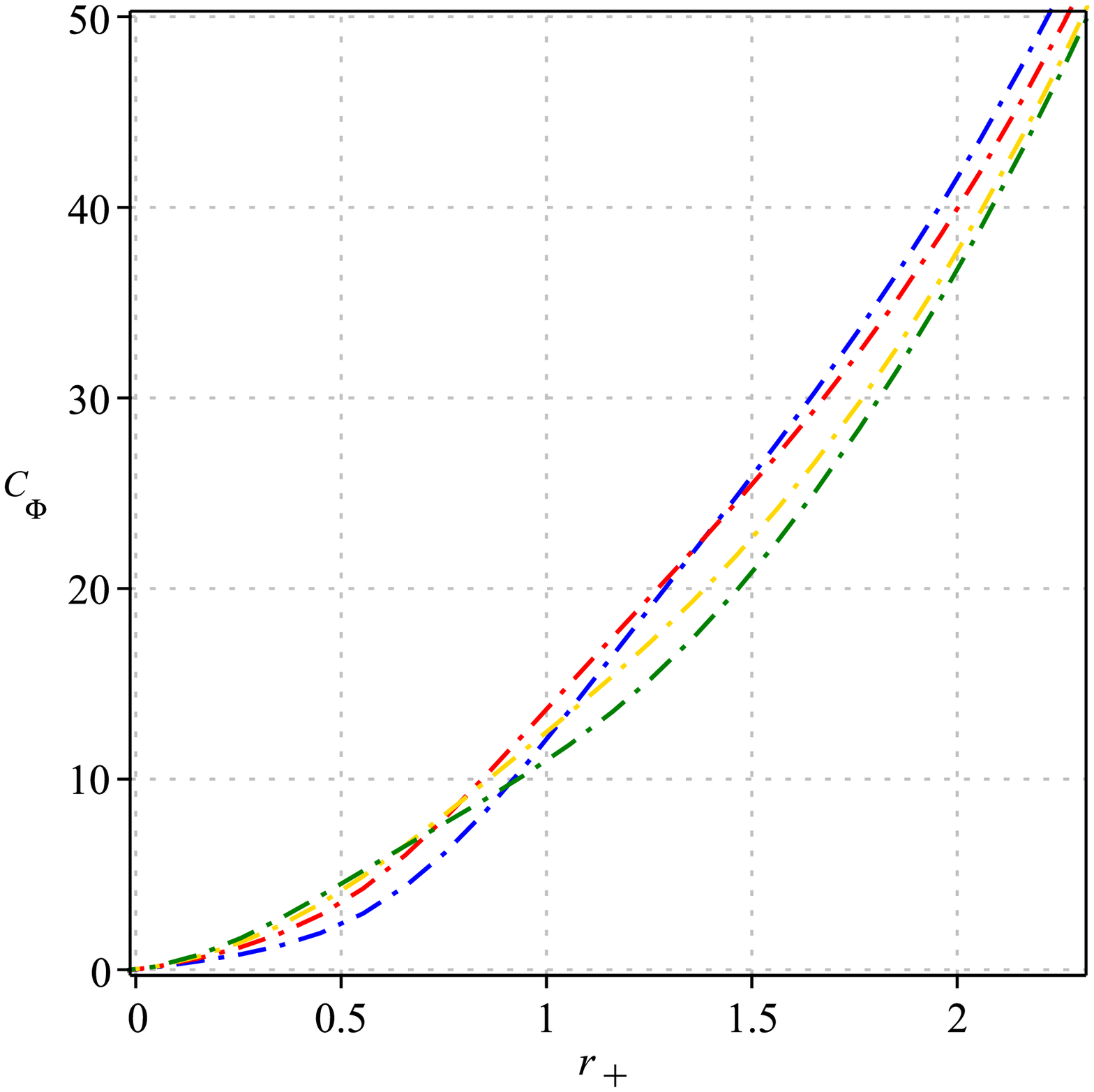}}\hfill
\subfloat[$\alpha=0.2$]{\includegraphics[width=.5\textwidth]{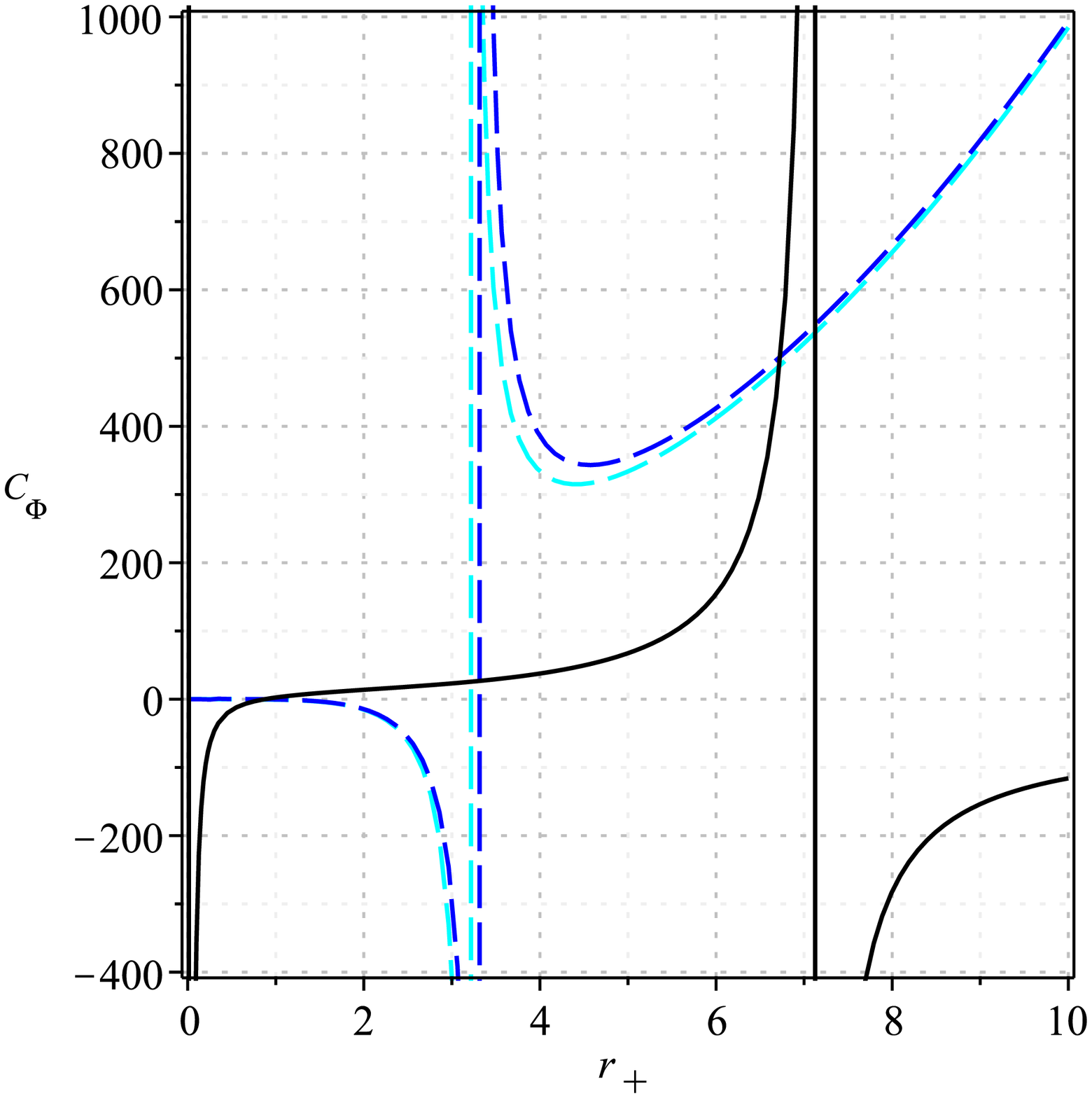}}\hfill
\caption{$\beta=0.05$ denoted by blue dash dot line, $\beta=0.5$ denoted by red dash dot line with, $\beta=1.0$ denoted by gold dash dot line and $\beta=1.5$ denoted by green dash dot line in EGB-ENLE. Solid black line denotes EGB-Maxwell, cyan dash line denotes GR-ENLE and blue dash line denotes $\beta=0.5$ in GR-Maxwell with   $Q=1$, $l=2$ and $M=5$.}\label{fig:7}
\end{figure}

\begin{figure}[H]
\centering
\subfloat[$\alpha=0.4$]{\includegraphics[width=.5\textwidth]{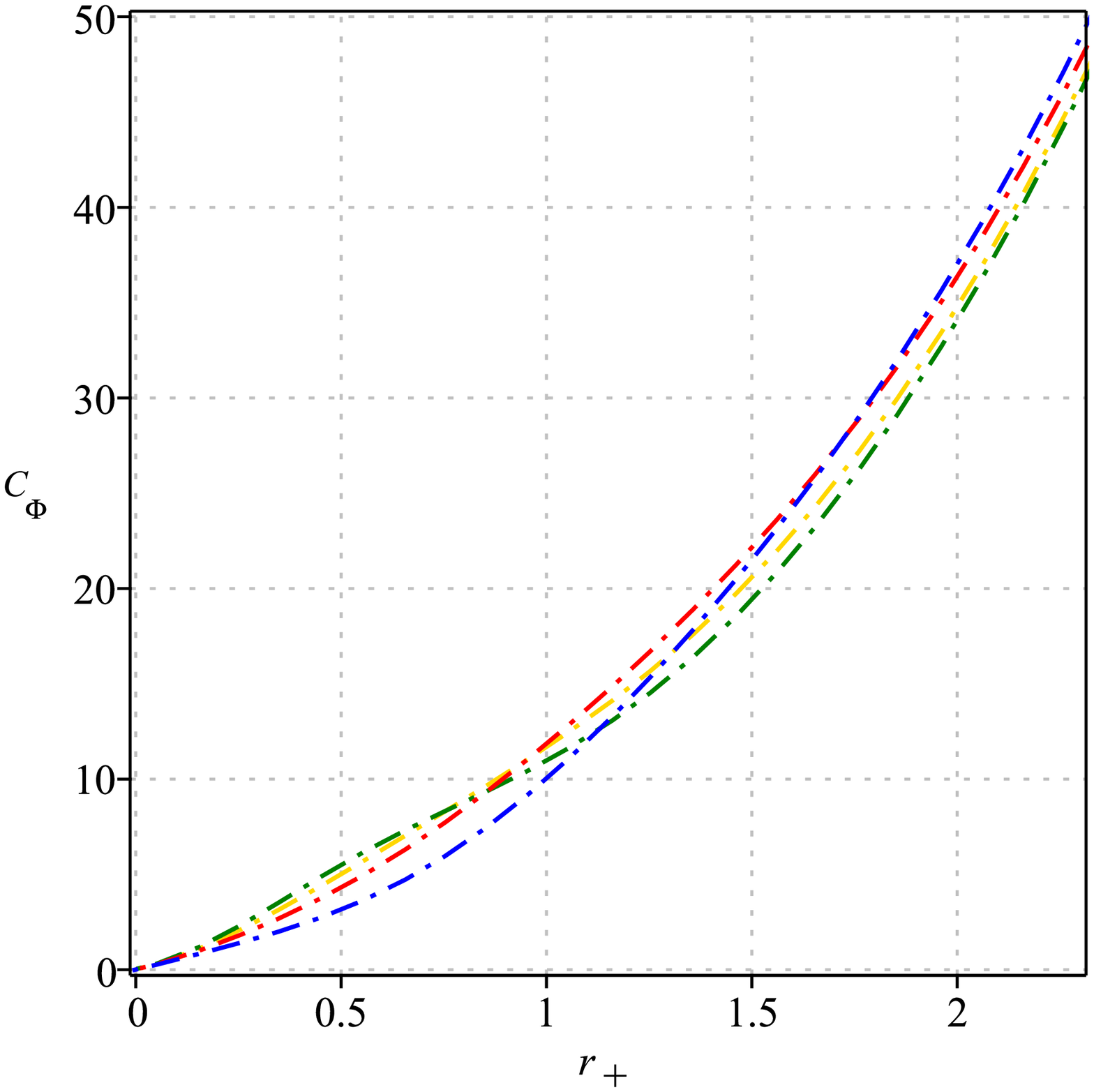}}\hfill
\subfloat[$\alpha=0.4$]{\includegraphics[width=.5\textwidth]{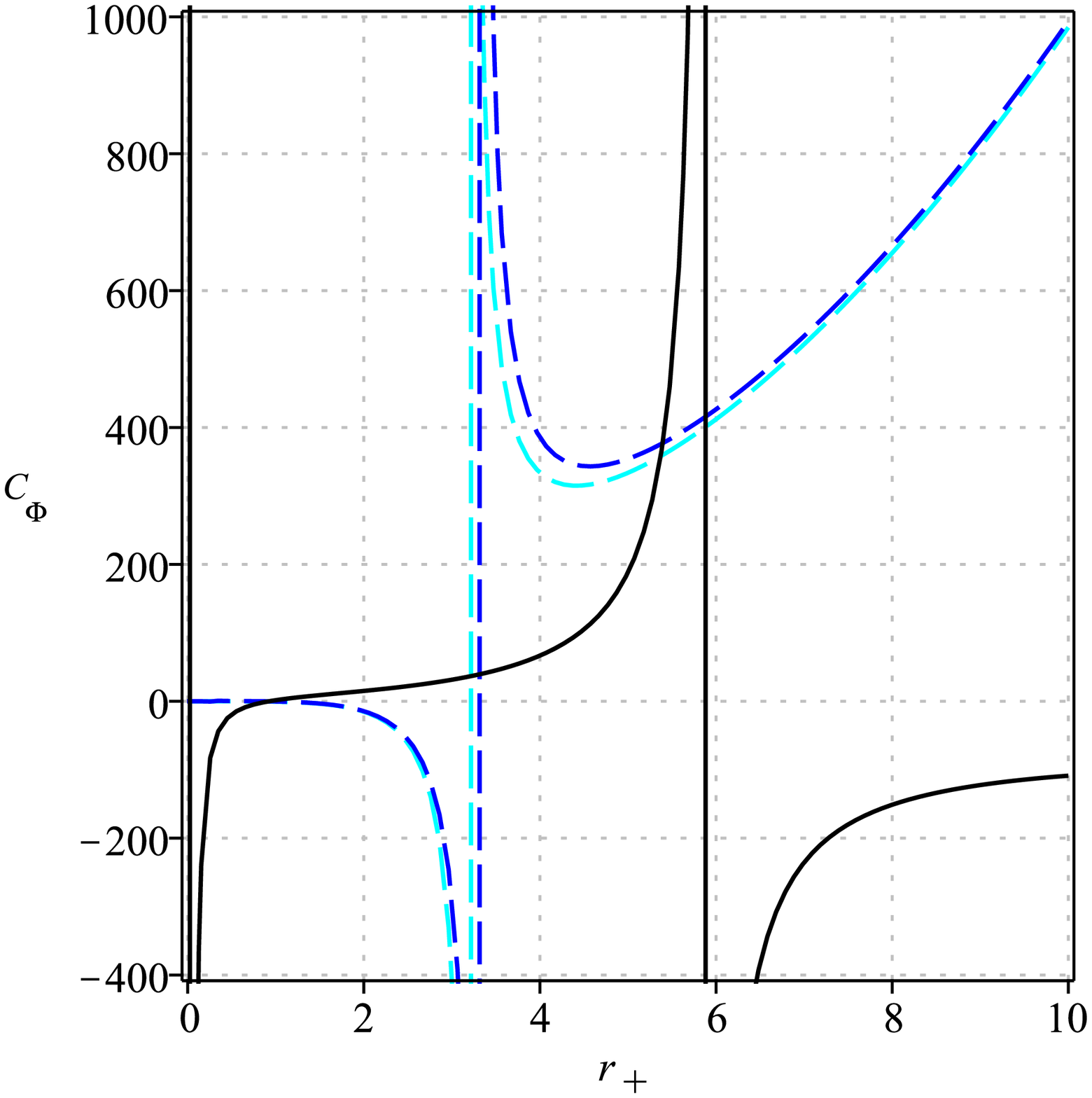}}\hfill
\caption{$\beta=0.05$ denoted by blue dash dot line, $\beta=0.5$ denoted by red dash dot line with, $\beta=1.0$ denoted by gold dash dot line and $\beta=1.5$ denoted by green dash dot line in EGB-ENLE. Solid black line denotes EGB-Maxwell, cyan dash line denotes GR-ENLE and blue dash line denotes $\beta=0.5$ in GR-Maxwell with   $Q=1$, $l=2$ and $M=5$.}\label{fig:8}
\end{figure}
The expression for black hole specific heat looks cumbersome, so we do not represent it here. The nature of the black hole specific heat for different values of $\beta$ and Gauss-Bonnet coupling parameters is shown in fig. \ref{fig:7} and fig. \ref{fig:8}. For all values of $\beta$ and Gauss-Binnet coupling parameters specific heat is positive and continuous, which indicates that black holes are locally stable. For a particular value of horizon radius(say, $r_{+}^{a}$), specific heat with $\beta=0.5, 1.0$ and $1.5$ intercept. Similarly, for a particular value of horizon radius(say, $r_{+}^{b}$ and $r_{+}^{b}>r_{+}^{a}$), specific heat with $\beta=0.05$ and $1.5$ intercept. The specific heat for  Einstein gravity with ENLE, $4D$ EGB  and Einstein gravity with Maxwell electrodynamics is discontinuous, i.e., a second order phase transition occurs for these black holes.

Next, we will find Helmholtz free energy of the black hole to study global stability
\begin{equation}\label{eq:3.11}
    F= M-T_{H}S, 
\end{equation}

using equations \eqref{eq:3.1}, \eqref{eq:3.3}, \eqref{eq:3.5} and \eqref{eq:3.7} we obtain
\begin{equation*}
F=-\frac{\ln  (r_{{+}} ) \alpha  r_{{+}}^{3} \beta^{2}}{2 r_{{+}}^{2}+4 \alpha}-\frac{\ln  (r_{{+}} ) \alpha  r_{{+}}}{r_{{+}}^{2}+2 \alpha}-\frac{3 \ln  (r_{{+}} ) \alpha  r_{{+}}^{3}}{(r_{{+}}^{2}+2 \alpha ) l^{2}}+\frac{\ln  (r_{{+}} ) \alpha^{2}}{(r_{{+}}^{2}+2 \alpha ) r_{{+}}}-\frac{r_{{+}}^{5} \beta^{2}}{8 r_{{+}}^{2}+16 \alpha}-\frac{r_{{+}}^{3}}{4 r_{{+}}^{2}+8 \alpha}-\frac{3 r_{{+}}^{5}}{4 (r_{{+}}^{2}+2 \alpha ) l^{2}}
\end{equation*}
\begin{equation*}
+\frac{r_{{+}} \alpha}{4 r_{{+}}^{2}+8 \alpha}+\frac{\alpha}{2 r_{{+}}}+\frac{r_{{+}}}{2}+\frac{\beta^{2} r_{{+}}^{3}}{12}+\frac{r_{{+}}^{3}}{2 l^{2}}+\frac{\sqrt{2\beta Q^3}  (2 L_W -1)}{6 \sqrt{\sqrt{x}L_W } }+\frac{r \beta Q   (4 \ln(r_+) \alpha + r_+^{2})}{\sqrt{L_W } (4 r_+^{2}+8 \alpha )}
\end{equation*}
\begin{equation}\label{eq:3.12}
+\frac{\beta^{2} L_W^{\frac{5}{4}} 2^{\frac{3}{2}}  (Q^{2}/{\beta^{2}})^{\frac{3}{4}} {}_{1}F_{1} \Bigl( 1;\frac{9}{4};\frac{L_W}{4}\Bigl) { e}^{-\frac{L_W}{4}}}{30}
\end{equation}

\begin{figure}[H]
\centering
\subfloat[$\alpha=0.2$]{\includegraphics[width=.5\textwidth]{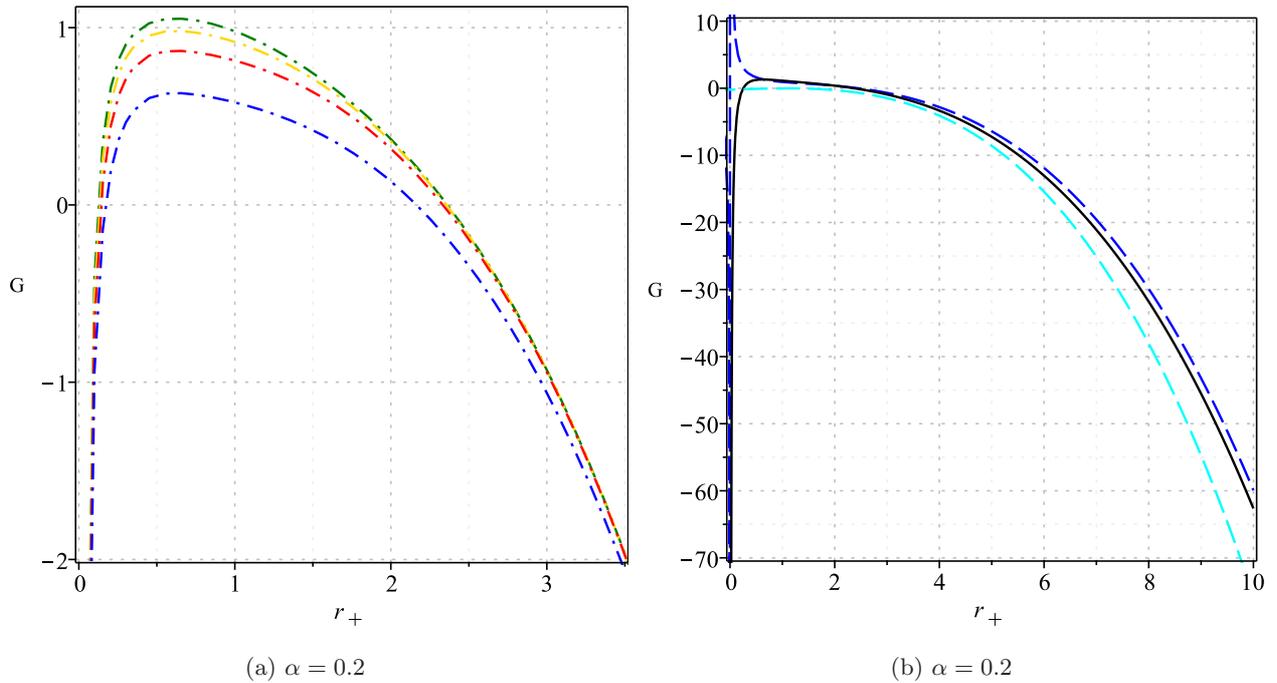}}\hfill
\subfloat[$\alpha=0.2$]{\includegraphics[width=.5\textwidth]{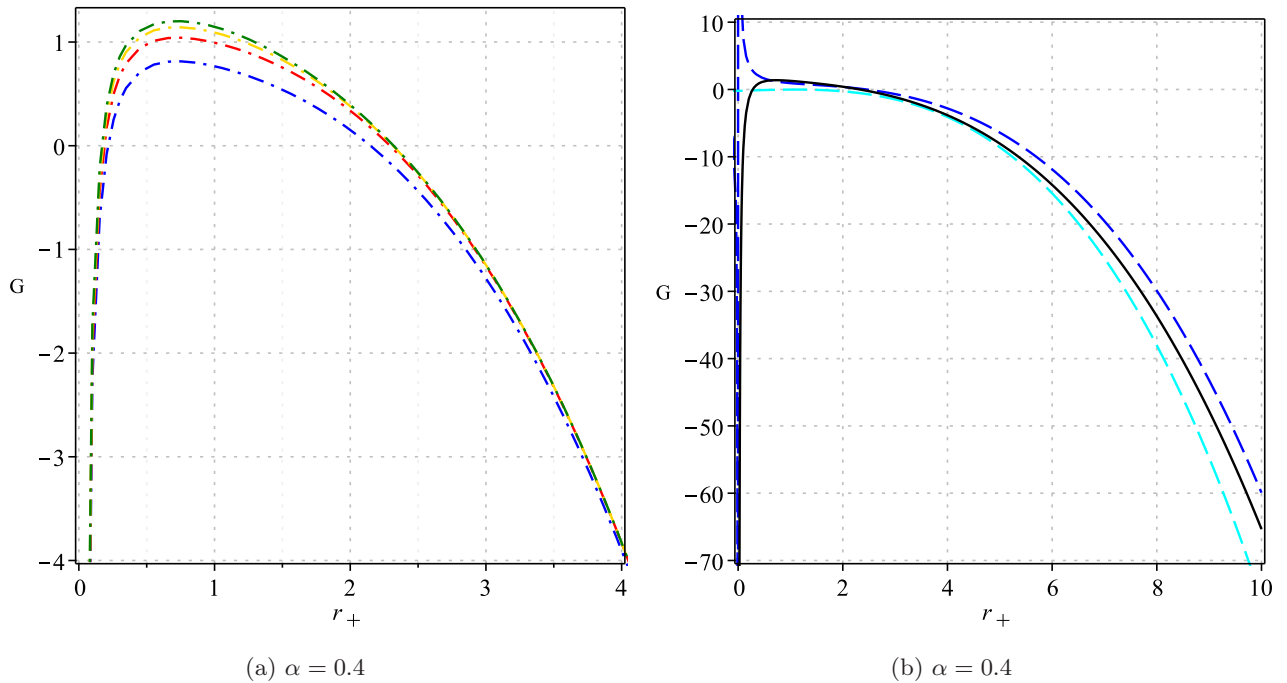}}\hfill
\caption{$\beta=0.05$ denoted by blue dash dot line, $\beta=0.5$ denoted by red dash dot line with, $\beta=1.0$ denoted by gold dash dot line and $\beta=1.5$ denoted by green dash dot line in EGB-ENLE. Solid black line denotes EGB-Maxwell, cyan dash line denotes GR-ENLE and blue dash line denotes $\beta=0.5$ in GR-Maxwell with   $Q=1$, $l=2$ and $M=5$.}\label{fig:9}
\end{figure}

\begin{figure}[H]
\centering
\subfloat[$\alpha=0.4$]{\includegraphics[width=.5\textwidth]{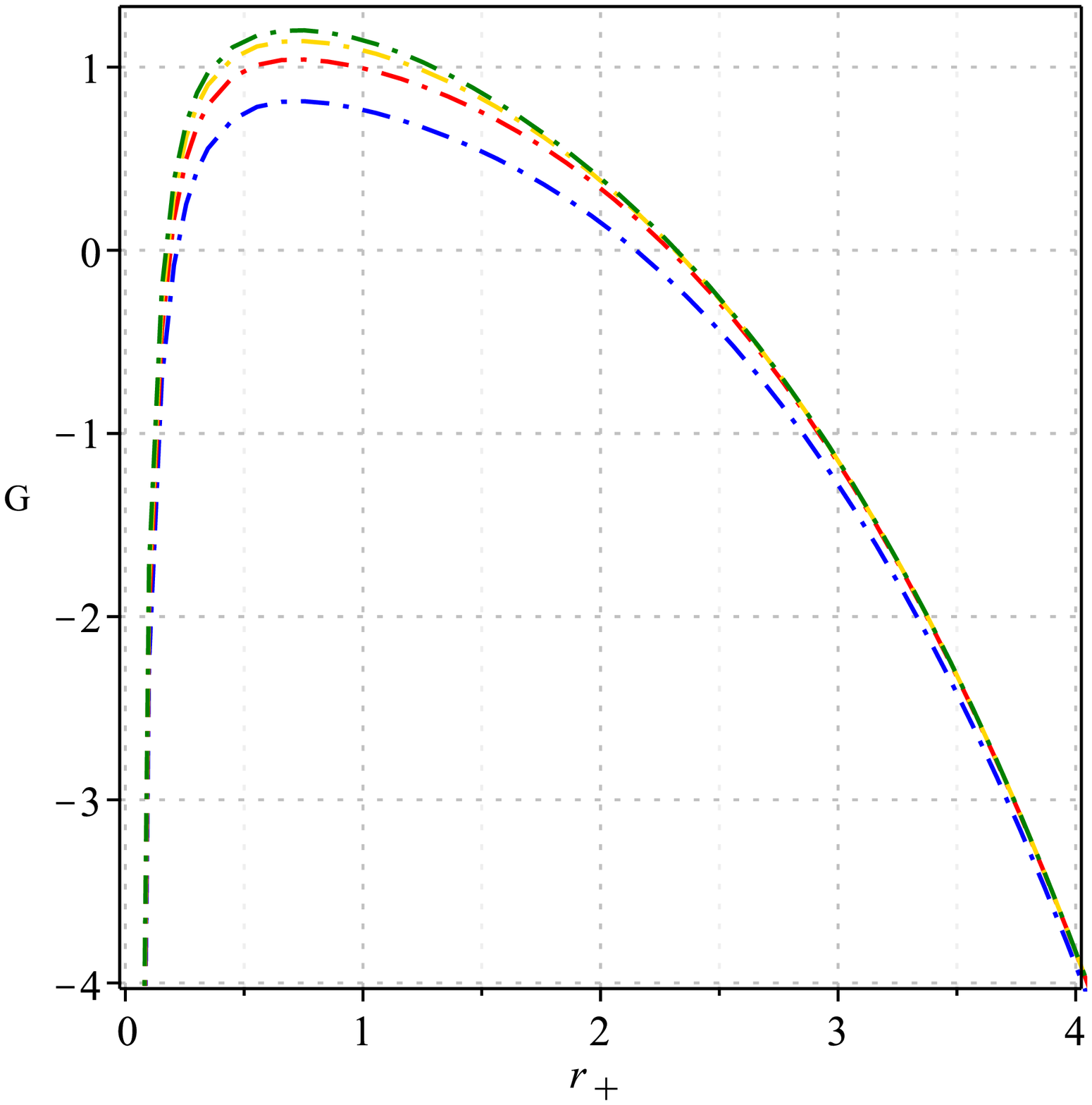}}\hfill
\subfloat[$\alpha=0.4$]{\includegraphics[width=.5\textwidth]{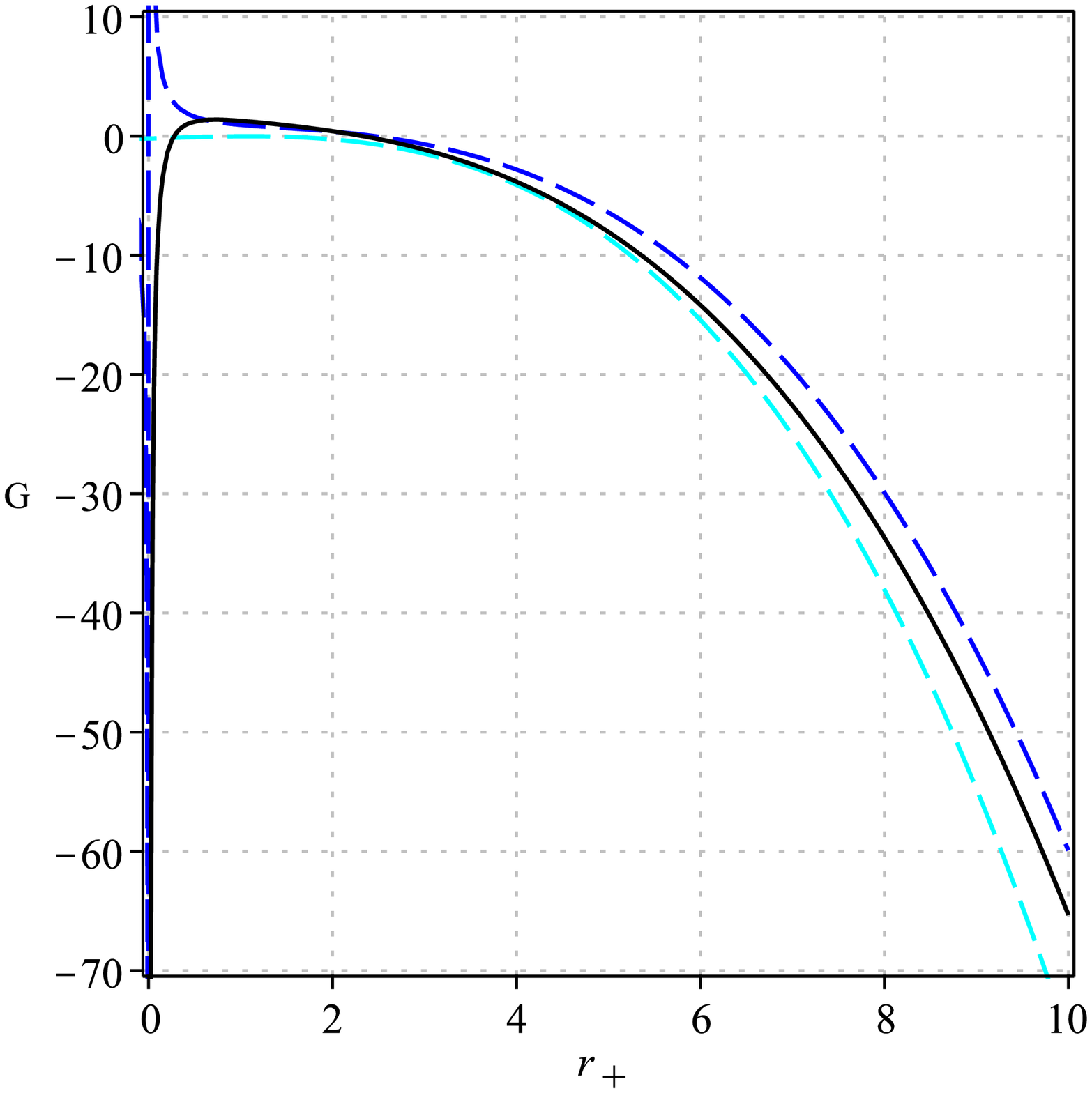}}\hfill
\caption{$\beta=0.05$ denoted by blue dash dot line, $\beta=0.5$ denoted by red dash dot line with, $\beta=1.0$ denoted by gold dash dot line and $\beta=1.5$ denoted by green dash dot line in EGB-ENLE. Solid black line denotes EGB-Maxwell, cyan dash line denotes GR-ENLE and blue dash line denotes $\beta=0.5$ in GR-Maxwell with   $Q=1$, $l=2$ and $M=5$.}\label{fig:10}
\end{figure}

In fig. \ref{fig:9} and fig. \ref{fig:10} Helmholtz free energy is depicted for different values of $\beta$ and Gauss-Bonnet coupling parameters. For two values of black hole horizon(say, $r_+^c$ and $r_+^c$ with $r_+^c>r_+^d$) Helmholtz free energy is zero. Helmholtz free energy is negative for $r_+<r_+^c$, which indicates that black holes with horizon radius $r<r_+^c$ are thermodynamically more stable. In the region between $r_+^c<r<r_+^d$, Helmholtz free energy takes positive and maximum values. Black hole with the radius $r_+^c<r<r_+^d$ are thermodynamically unstable. If we increase the $\beta$ values then positive and maximum parts are increased. After crossing the point $r=r_+^d$, Helmholtz free energy once again takes negative value, i.e. black hole with the radius $r>r_+^d$ are thermodynamically more stable.

\section{Conclusion}\label{sec:4}
In this work, we obtained an electrically charged black hole solution in $4D$ Einstein-Gauss-Bonnet gravity coupled to exponential nonlinear electrodynamics. The horizon structure of the black hole is depicted in fig. \ref{fig:1} and fig. \ref{fig:2}. From the plot, it is clear that the black hole has two horizons. One is inner horizon and another is outer horizon, outer horizon is known as the black hole horizon. The mass and Hawking temperature of the black hole were analysed for different values of nonlinear electrodynamics and Gauss-Bonnet coupling parameters. The thermodynamics of the black hole studied in extended phase space, where cosmological constant played the role of pressure. Taking the Gauss-Bonnet coupling parameter and nonlinear electrodynamics parameters as variables, we obtained the first law of black hole thermodynamics. The specific heat of the black hole is computed, it is positive and an increasing function of horizon radius, which indicates that black holes are locally stable. Furthermore, we compared the thermodynamics of $4D$ EGB gravity(ENLD) with Einstein gravity(ENLD), $4D$ EGB(Maxwell) and  Einstein gravity(Maxwell). Finally, The Helmholtz free energy of nonlinearly charged $4D$ Einstein-Gauss-Bonnet black holes are studied, which shows that very small and larger sized black holes are globally more stable. The intermediate sized black hole are not thermodynamically stable.

\printbibliography
\end{document}